\documentclass{jfm}
\usepackage{graphicx,epsfig}
\usepackage{natbib}
\usepackage{mathrsfs}
\usepackage{amsmath}
\usepackage{amsfonts}
\usepackage{amssymb}
\usepackage{color}
\usepackage{caption}
\usepackage{float}
\usepackage{subcaption}
\usepackage{tikz,pgfplots}
\usepackage[font=small,labelfont=bf]{caption}
\usepackage{newtxtext}
\usepackage{newtxmath}
\usepackage{hyperref}
\usepackage{enumerate}

\ifCUPmtlplainloaded \else
  \checkfont{eurm10}
  \iffontfound
    \IfFileExists{upmath.sty}
      {\typeout{^^JFound AMS Euler Roman fonts on the system,
                   using the 'upmath' package.^^J}%
       \usepackage{upmath}}
      {\typeout{^^JFound AMS Euler Roman fonts on the system, but you
                   dont seem to have the}%
       \typeout{'upmath' package installed. JFM.cls can take advantage
                 of these fonts,^^Jif you use 'upmath' package.^^J}%
      }
  \else
  \fi
\fi

\ifCUPmtlplainloaded \else
  \checkfont{msam10}
  \iffontfound
    \IfFileExists{amssymb.sty}
      {\typeout{^^JFound AMS Symbol fonts on the system, using the
                'amssymb' package.^^J}%
       \usepackage{amssymb}%
         \let\leq=\leqslant
         
      }{}
  \fi
\fi

\ifCUPmtlplainloaded \else
  \IfFileExists{amsbsy.sty}
    {\typeout{^^JFound the 'amsbsy' package on the system, using it.^^J}%
     \usepackage{amsbsy}}
    {}
\fi

\newsavebox{\astrutbox}
\sbox{\astrutbox}{\rule[-5pt]{0pt}{20pt}}

\newcommand\beq{\begin{equation}}
\newcommand\eeq{\end{equation}}
\newcommand\beqa{\begin{eqnarray}}
\newcommand\eeqa{\end{eqnarray}}

\newcommand{\bnab}{\mbox{\boldmath$\nabla$}}

\newcommand{\eps}{\varepsilon}
\newcommand{\del}{\delta}
\def\bnab{\mbox{\boldmath $ \nabla$}}

\def\half {{\textstyle{1 \over 2}}}

\newcommand{\bhx}{\mathbf{\hat{x}}}
\newcommand{\bhy}{\mathbf{\hat{y}}}

\newcommand{\bu}{\mathbf{u}}
\newcommand{\bv}{\mathbf{v}}
\newcommand{\btau}{\mbox{\boldmath{$\tau$}}}
\newcommand{\bTau}{\mbox{\boldmath{${\cal T}$}}}
\newcommand{\bU}{\mathbf{U}}

\def\<{\langle}
\def\>{\rangle}

\newcommand{\hD}{\hat{D}}
\newcommand{\htau}{\hat{\tau}}
\newcommand{\lam}{\Lambda}
\newcommand{\Us}{U_{*}^{'}}
\newcommand{\Om}{\Omega}

\newcommand{\hu}{\hat{u}}
\newcommand{\hv}{\hat{v}}
\newcommand{\hp}{\hat{p}}
\newcommand{\htxx}{\hat{\tau}_{11}}
\newcommand{\htxy}{\hat{\tau}_{12}}
\newcommand{\htyy}{\hat{\tau}_{22}}
\newcommand{\hk}{\hat{k}}
\newcommand{\hW}{\hat{W}}
\newcommand{\ha}{\hat{a}}
\newcommand{\hY}{\hat{Y}}

\title[]{Asymptotics of the centre mode instability in viscoelastic channel flow: with and without inertia}

\author[R. R. Kerswell \& J. Page]%
{ Rich R. Kerswell$^1$\thanks{r.r.kerswell@damtp.cam.ac.uk}\ns and
Jacob Page$^2$\thanks{jacob.page@ed.ac.uk}}

\affiliation{$^1$DAMTP, Centre for Mathematical Sciences, University of Cambridge, CB3 0WA, UK.\\
$^2$School of Mathematics, University of Edinburgh, EH9 3FD, UK}

\date{?; revised ?; accepted ?.}

\begin{document}

\maketitle

\begin{abstract}

Motivated by the recent numerical results of Khalid et al.,  {\em Phys. Rev. Lett.}, {\bf 127} 134502 (2021), we consider the large-Weissenberg-number ($W$) asymptotics of
the centre mode instability in inertialess viscoelastic channel flow. The instability is of the critical layer type in the distinguished ultra-dilute limit where $W(1-\beta)=O(1)$ as $W \rightarrow \infty$ ($\beta$ is the ratio of solvent-to-total viscosity). In contrast to centre modes in the Orr-Sommerfeld equation, $1-c=O(1)$ as $W \rightarrow \infty$ where $c$ is the phase speed normalised by the centreline speed as a central `outer' region is always needed to adjust the non-zero cross-stream velocity at the critical layer down to zero at the centreline. The critical layer acts as a pair of intense `bellows' which blows the flow streamlines apart locally and then sucks them back together again. This compression/rarefaction amplifies the streamwise-normal polymer stress which in turn drives the streamwise flow through local polymer stresses at the critical layer. The streamwise flow energises the cross-stream flow via continuity  which in turn intensifies the critical layer to close the cycle. We also treat the large-Reynolds-number ($Re$) asymptotic structure of the upper (where $1-c=O(Re^{-2/3})$) and lower branches of the $Re$-$W$ neutral curve confirming the inferred scalings from previous numerical computations.  Finally, we argue that the viscoelastic centre mode instability was actually first found in viscoelastic Kolmogorov flow by Boffetta et al.,  {\em J. Fluid Mech.} {\bf 523}, 161-170 (2005).

\end{abstract}


%
%
\section{Introduction}

It is well known that the addition of  long-chain polymers to a Newtonian fluid  introduces elasticity  which can give rise to fascinating new `viscoelastic' flow phenomena. Prime examples of this are a new form of spatiotemporal chaos - dubbed `elastic' turbulence (ET) \citep{groisman00} - which exists in inertialess curvilinear flows and `elasto-inertial' turbulence (EIT) \citep{samanta13} which can occur in 2D rectilinear flows \citep{Sid18} where inertia and elasticity balance each other. While ET is assumed triggered by a linear `hoop stress' instability of curved streamlines \cite{Larson90,Shaqfeh96}, the origin of EIT remains unclear \citep{Datta22, Dubief23} as does any possible relationship to ET. 

The breakdown of viscoelastically-modified Tollmien-Schlichting modes has been suggested as a cause of EIT \citep{shekar19,shekar21} at least at high Reynolds number, $Re$ and low Weissenberg number, $W$. At low $Re$ and high $W$, however, the recent discovery of a new linear instability of rectilinear viscoelastic shear flow seems  more viable \citep{Garg18, Chaudhary21, Khalid21a}. This instability occurs at higher $W$ than generally associated with EIT  but has been shown to be subcritical \citep{Page20,Wan21,Buza22a,Buza22b}. In particular, travelling wave solutions, which have a distinctive `arrowhead' structure, originating from the neutral curve reach down in $W$ to where EIT exists in parameter space \citep{Page20, Buza22b, Dubief22}. This instability is of centre-mode type being localised either at the centre of a pipe \citep{Garg18,Chaudhary21} or midplane of a  channel \cite{Khalid21a,Khalid21b} but is notably absent in plane Couette flow \citep{Garg18}. Perhaps most intriguingly, the instability can be traced down to $Re=0$ in channel flow \citep{Khalid21b} in the ultra-dilute limit of the solvent-to-total viscosity ratio approaching 1 while a minimum $Re \approx 63$ exists in pipe flow \citep{Chaudhary21}. Subsequently, travelling wave solutions have been numerically computed in 2 dimensions and at $Re=0$ \citep{Buza22b,Morozov22}  and their instability examined \citep{Lellep23a,Lellep23b}.

Apart from numerically-inferred scaling relationships \citep{Garg18,Chaudhary21,Khalid21a,Khalid21b}, the only work to unpick the asymptotic structure of the centre-mode instability is that of \cite{Dong22} in the pipe flow. They identify the asymptotic structure on the upper branch of the neutral curve  characterised by $W \sim Re^{1/3}$ as $Re \rightarrow \infty$ and consider the long wavelength limit but stop short of treating the lower branch of the neutral curve. Here we do both for the channel and go further to examine the inertialess regime in channel flow which is absent in pipe flow. Unravelling the $Re=0$ situation asymptotically is actually our  main motivation here as it differs fundamentally from all the classical Orr-Sommerfeld work performed for Newtonian shear flows \citep{DrazinReid}. In particular, the regularizing feature of the critical layer formed (e.g. figure 3 of \cite{Khalid21b} and figure \ref{eigenfunction} below) is the presence of elastic relaxation rather than viscosity and the `outer' relaxation-free solutions satisfy a fourth order differential equation rather than the classical, inviscid, second order Rayleigh equation in Newtonian flows. This means that matching conditions across the critical layer need to be  sought down to the third order derivative in the cross-stream velocity (or streamfunction) and, due to a logarithmic singularity in the first order derivative,  computations need to go beyond double precision accuracy to achieve a convincing correspondence between numerical results  and the asymptotic predictions; see Table \ref{Table4}. A particularly interesting feature of this viscoelastic centre mode instability is the critical layer does {\em not} approach the midplane as $W \rightarrow \infty$, that is, the phase speed of the instability approaches a non-trivial value very close to but distinct from 1, the maximum speed of the base flow. Ultimately, though,  the point of the asymptotic analysis is to identify the mechanism of the instability and to understand, if possible, why it does not manifest in plane Couette flow.

\begin{figure}
\centering
\scalebox{0.37}[0.37]{\includegraphics{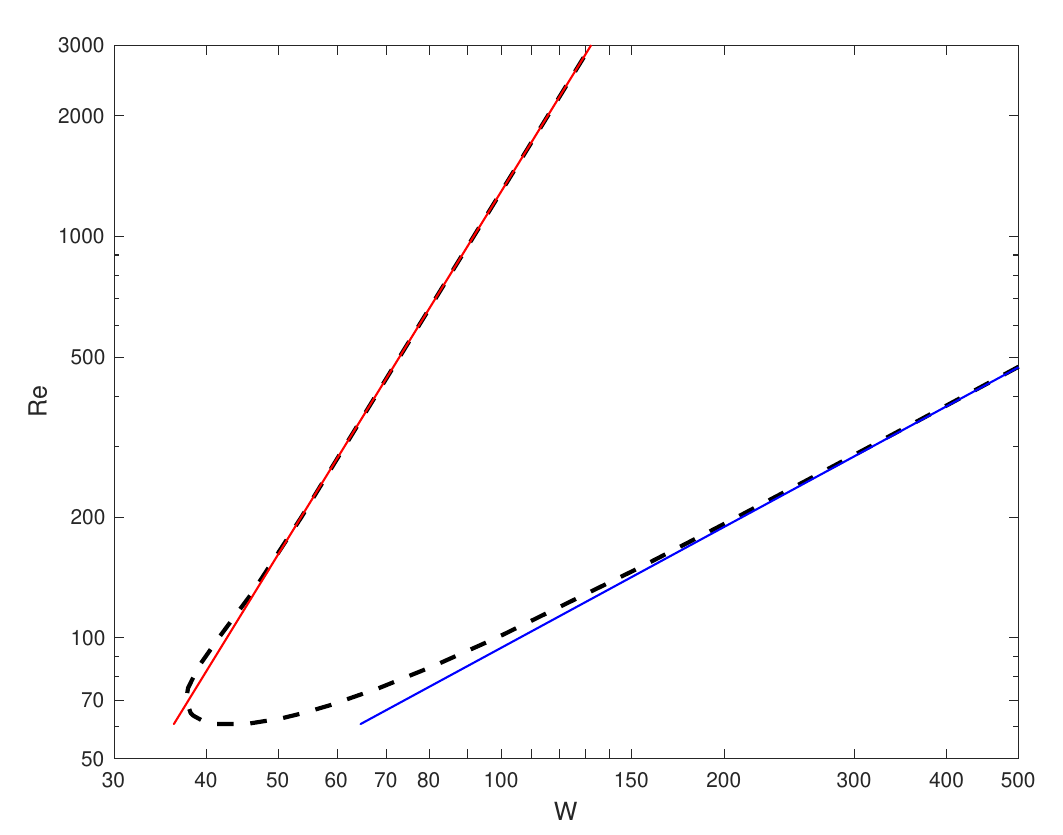}}
\scalebox{0.37}[0.37]{\includegraphics{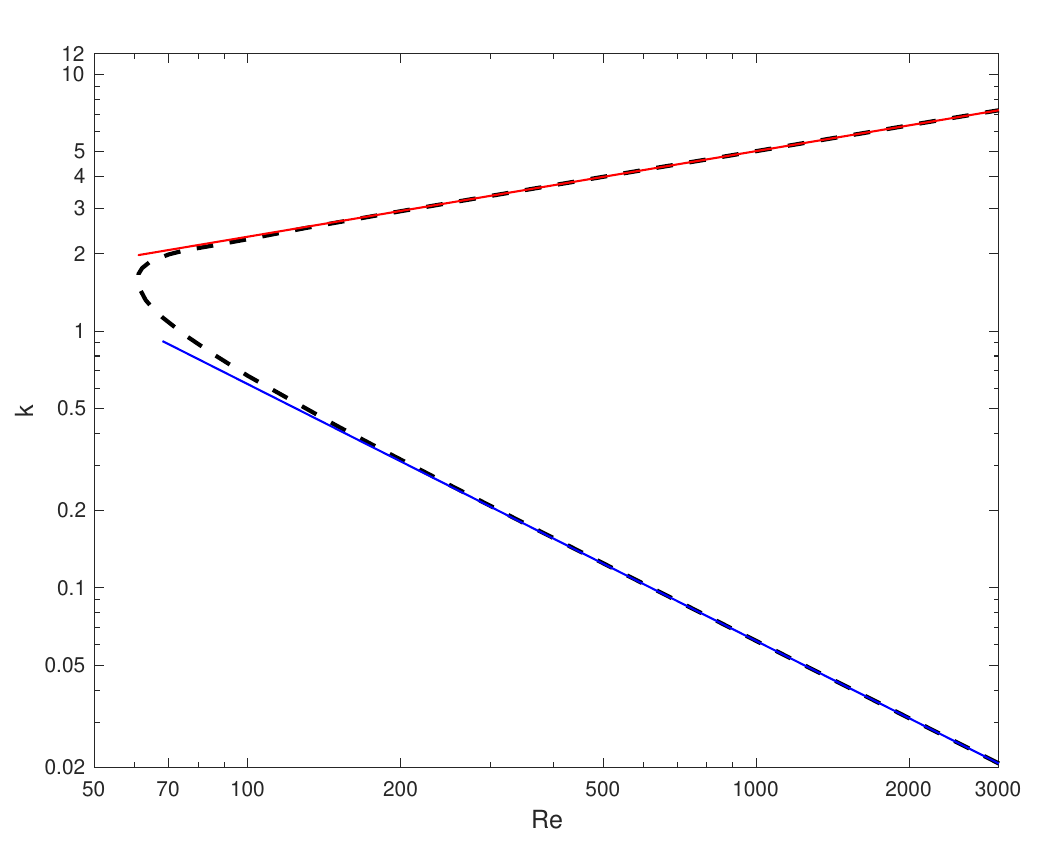}}\\
\scalebox{0.37}[0.37]{\includegraphics{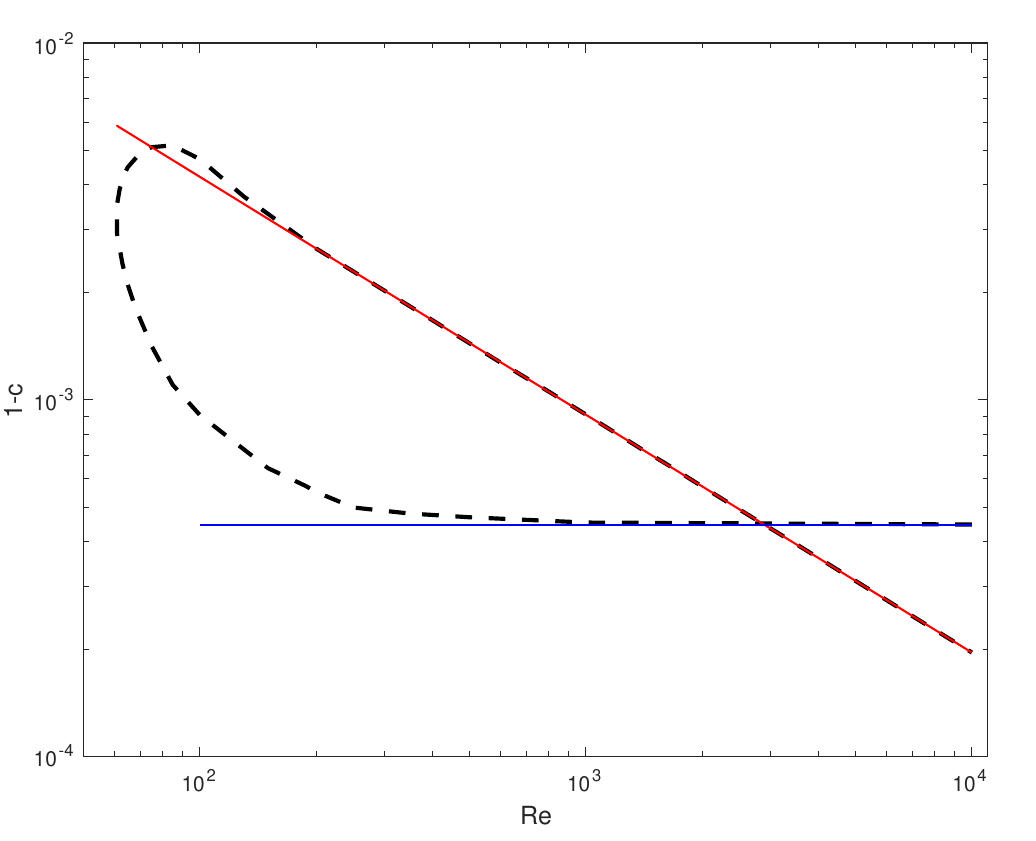}}
\scalebox{0.37}[0.37]{\includegraphics{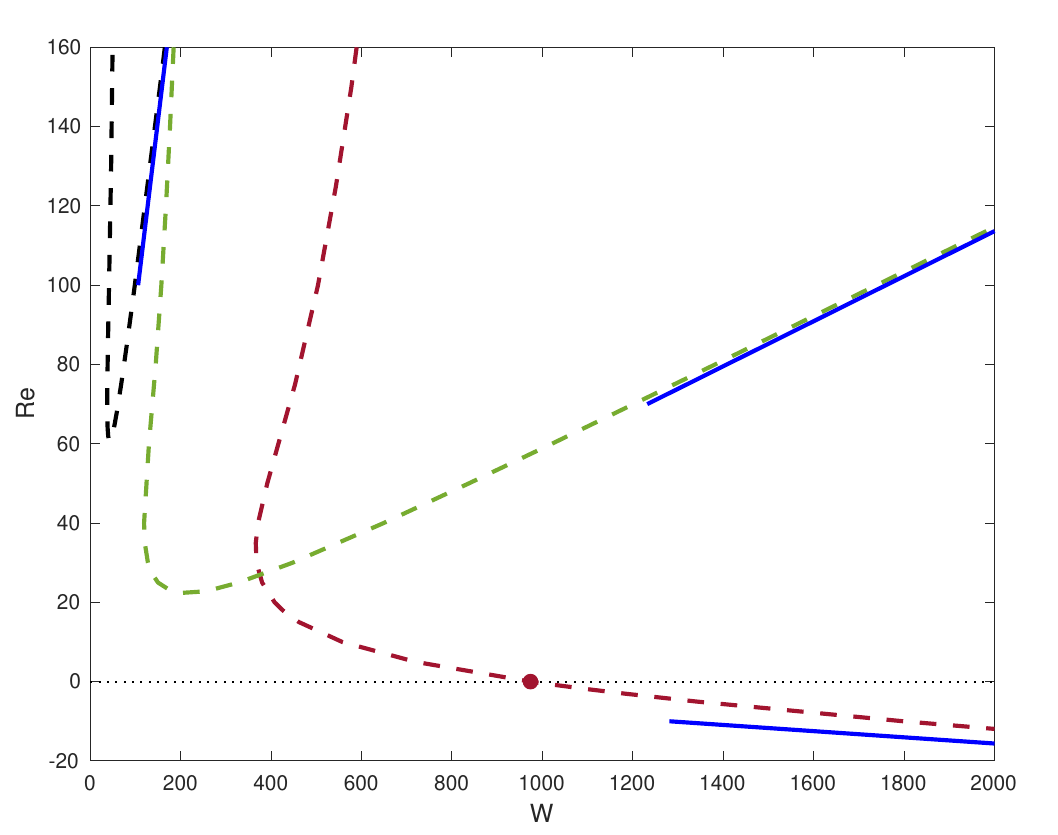}}
\caption{\label{ub_lb_curves} The centre mode neutral curve (black dashed line) for $\beta=0.9$ and the asymptotic predictions (upper/lower branch - red/blue solid lines) for $W$ (top left), $k$ (top right) and $1-c$ (bottom left). The bottom right plot shows how the lower branch of the neutral curve rotates around and crosses the inertialess limit of $Re=0$ as $\beta$ increases from $\beta=0.9$ (black again) through $\beta=0.98$ (green) to $\beta=0.994$ (dark red). The solid blue lines indicate the asymptotic prediction for the various lower branches (the $\beta=0.994$ prediction is reached for $Re \rightarrow -\infty$; note only $Re=O(10)$ is shown here). The dark red dot at $(W,Re)=(973.8,0)$ is the lowest point reached by the neutral curve for $Re=0$ for an Oldroyd-B fluid (\cite{Khalid21b}). The $W$-$Re$ neutral curve in the top left of this  figure is the channel flow equivalent of the pipe flow curve shown on the left in figure 1 of \cite{Dong22}.} 
\end{figure}

%
%
The plan of this paper is first to  introduce  the channel flow problem in \S 2 and the viscoelastic model (Oldroyd-B) used by \cite{Khalid21a,Khalid21b}. The first results section, \S 3, then examines the large $Re$-asymptotics of the upper (\S 3.1) and lower branches (\S 3.2) of the neutral curve in the $Re$-$W$ plane for fixed $\beta$: see figure \ref{ub_lb_curves}. Reduced eigenvalue problems based only on O(1) quantities (relative to $Re$) can be straightforwardly derived for both upper and lower branches. Interestingly, if $\beta \gtrsim 0.9905$, \cite{Khalid21b} showed that the lower branch crosses the $Re=0$ axis and the appropriate (mathematical) limit is then $Re \rightarrow -\infty$. Figure \ref{lb_efns} indicates that nothing mathematically untoward happens as the neutral curve swings around from pointing at $Re \rightarrow \infty$ to $Re \rightarrow -\infty$ although, of course, negative $Re$ makes little physical sense. The special case of $Re=0$ or vanishing inertia, however, does and the asymptotics as $W \rightarrow \infty$ is studied in \S4. The work of \cite{Khalid21b} has already indicated that the appropriate distinquished limit is that in which  $\beta$ simultaneously approaches 1 such that $W(1-\beta)$ stays finite. Section 5 goes on to use the asymptotic solution to discuss the mechanics of the inertialess instability
 and \S 6 describes some numerical experiments to understand how the instability responds to the problem becoming a bit more plane-Couette like. A brief \S7 presents evidence that the centre mode instability was actually found first in viscoelastic Kolmogorov flow \citep{Boffetta05}  before a final discussion follows in \S8.

%
%
\section{Formulation}
%

We consider pressure-driven, incompressible channel flow between two walls $y=\pm h$ in the $x$-direction. Using the half-channel height, $h$, and the base centreline speed $U_{max}$ to non-dimensionalize the problem,  the governing equations become
\beq
Re \left( \frac{\partial \bu}{\partial t} + \bu \cdot \bnab \bu \right)=- \nabla {\cal P} +\beta \nabla^2 \bu +\bnab \cdot \bTau,
\eeq
\beq
\nabla \cdot \bu =0,
\eeq
\beq 
\frac{\partial \bTau}{\partial t}+\bu \cdot \bnab \bTau-2 \,{\rm sym}(\bTau.\bnab  \bu) =-\frac{1}{W}\bTau+\frac{1-\beta}{W} (\,\bnab\bu +\bnab \bu^T\,)
\eeq
where $\bu$ is the velocity field, ${\cal P}$ the pressure and $\bTau$ the polymer stress following \cite{Khalid21a}.
Here an  Oldroyd-B fluid has been assumed so
\beq
\bTau= \frac{1-\beta}{W}({\bf C}-{\bf I})
\eeq
where ${\bf C}$ is the conformation tensor. The parameters of the problem are  the Reynolds number, Weissenberg number and the 
solvent-to-total viscosity ratio,
\beq
Re:= \frac{U_{max}h}{\nu},\quad W:= \frac{\lambda U_{max}}{h} \quad \& \quad \beta :=\frac{\nu_s}{\nu}
\eeq
respectively, where $\lambda$ is the microstructural relaxation time, $\nu_s$ is the solvent kinematic viscosity and $\nu$ is the total kinematic viscosity (following \cite{Khalid21a,Khalid21b}). The scaling of the pressure has been done in anticipation of  setting $Re=0$ in \S 4.\\

The 1-dimensional base state is 
\beq
\bU=U(y) \bhx:=(1-y^2) \bhx, \quad \bnab P= -2 \bhx, 
\eeq
\beq
{\bf T}= 
\left[ \begin{array}{ll}
T_{11} & T_{12} \\
T_{12} &  T_{22}    
\end{array}
 \right]
=
(1-\beta) 
\left[ \begin{array}{ll}
2W U'^{2} & U' \\
U'              & 0 
\end{array}
 \right]
\label{base}
\eeq
where $U':=dU/dy$ and henceforth, the analysis is entirely 2-dimensional. The linearized equations for small perturbations
\beq
\bv=u \bhx+v \bhy:=\bu-\bU, \quad p:={\cal P}-P \quad \& \quad \btau= \left[ \begin{array}{ll}
\tau_{11} & \tau_{12}\\
\tau_{12}              & \tau_{22}
\end{array}
 \right]:=\bTau-{\bf T}
\eeq
which are all assumed proportional to $e^{ik(x-ct)}$ where $k \in \mathbb{R} $ is a real wavenumber but the frequency $c=c_r+ic_i \in \mathbb{C}$ can be complex ($c_i>0 $ indicates instability; $c_r, c_i \in \mathbb{R}$), are the momentum and incompressibility equations,
\begin{align}
Re \left[ \, ik (U-c)u+U'v \,\right] & = -ikp+\beta (D^2-k^2) u + ik \tau_{11}+D \tau_{12}     \label{mom_x}\\
Re \left[ \, ik(U-c) v    \, \right] & =-Dp+\beta (D^2-k^2) v + ik \tau_{12}+D \tau_{22} ,     \label{mom_y}\\
iku+Dv &=0 \label{incompressible}
\end{align}
for the velocity field and
\begin{align}
\left[\frac{1}{W}+ik(U-c) \right] \tau_{11}  &= -v DT_{11} +2ikT_{11}u+2T_{12}Du+2U' \tau_{12} \hspace{-0.75cm}&+\frac{2i k (1-\beta)}{W} u,               
\label{eqn_tau_11}\\
\left[\frac{1}{W}+ik(U-c) \right] \tau_{12}  &= -v DT_{12} +ik T_{11}v +U'  \tau_{22}&+\frac{1-\beta}{W}(Du+ikv),
\label{eqn_tau_12}\\
\left[\frac{1}{W}+ik(U-c) \right] \tau_{22}  &= \,\,\,\,\,2ikT_{12} v &+ \frac{2 (1-\beta)}{W} Dv
\label{eqn_tau_22}
\end{align}
for the polymer field where $D:=d/dy$, $T_{11}=2 \lam U'^2$, $T_{12}=\lam U'/W$ and $T_{22}=0$  and $\lam:=W(1-\beta)$ (see (\ref{base})) in preparation for \S 4. The pressure $p$ can be eliminated between (\ref{mom_x}) and (\ref{mom_y}) to produce the vorticity equation
\begin{align}
\beta (D^2-k^2)^2 v= -k^2 D(\tau_{11}-\tau_{22})&+ik(D^2+k^2)\tau_{12} \nonumber \\
&\hspace{1cm}+ik Re\left[ (U-c)(D^2-k^2)v-U'' v\right]
\label{vorticity}
\end{align}
This equation is good for (asymptotic) analysis but not for a numerical solution where discretizing two 2nd order equations rather than one 4th order equation is far better conditioned process.
%
%
%

%
%

\section{$Re \rightarrow \infty$ Asymptotics for Channel Flow}

A natural starting point for examining the centre mode instability is to consider the neutral curve in the $Re$-$W$ plane for fixed $\beta$ (e.g. figure 2 of \cite{Page20}, figure \ref{ub_lb_curves} here for $\beta \in \{0.9, 0.98, 0.994 \}$ and figure 1 in \cite{Dong22} for pipe flow). The upper and lower branches of this neutral curve have $|Re| \rightarrow \infty$ limits which are now explored.

%
%
\subsection{Upper branch in $Re$ vs $W$ plane at fixed $\beta$ \label{upper} } 

\begin{figure}
\centering
\scalebox{0.38}[0.38]{\includegraphics{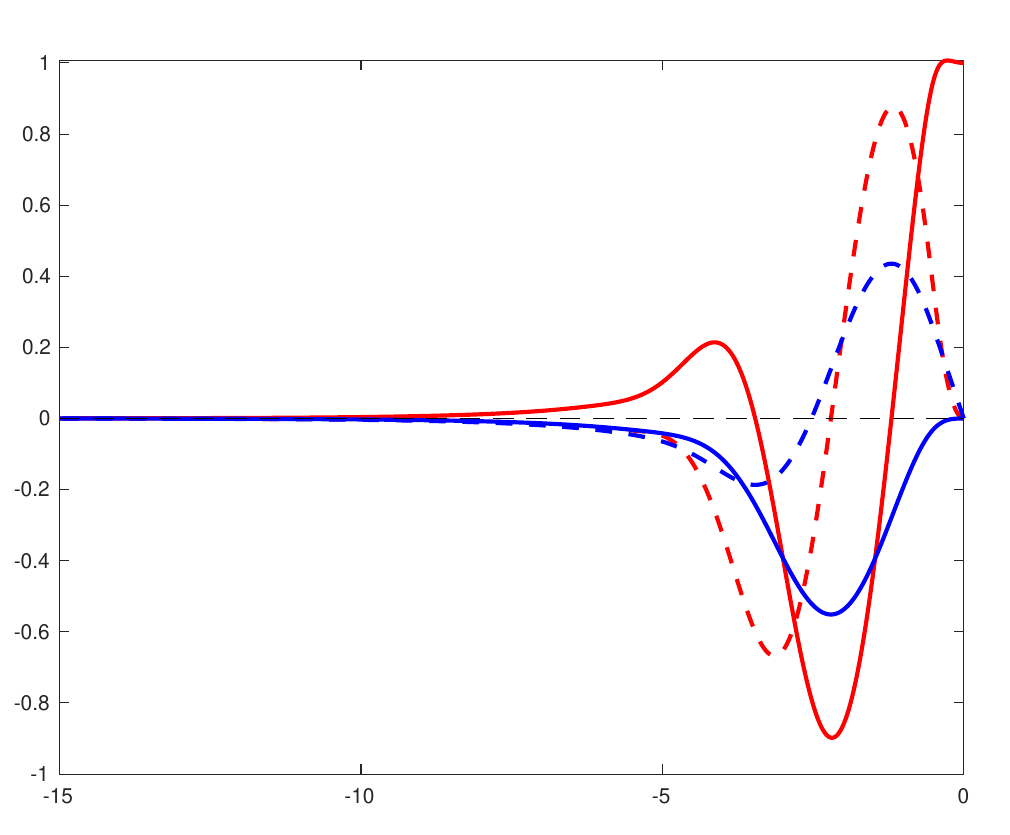}}
\scalebox{0.38}[0.38]{\includegraphics{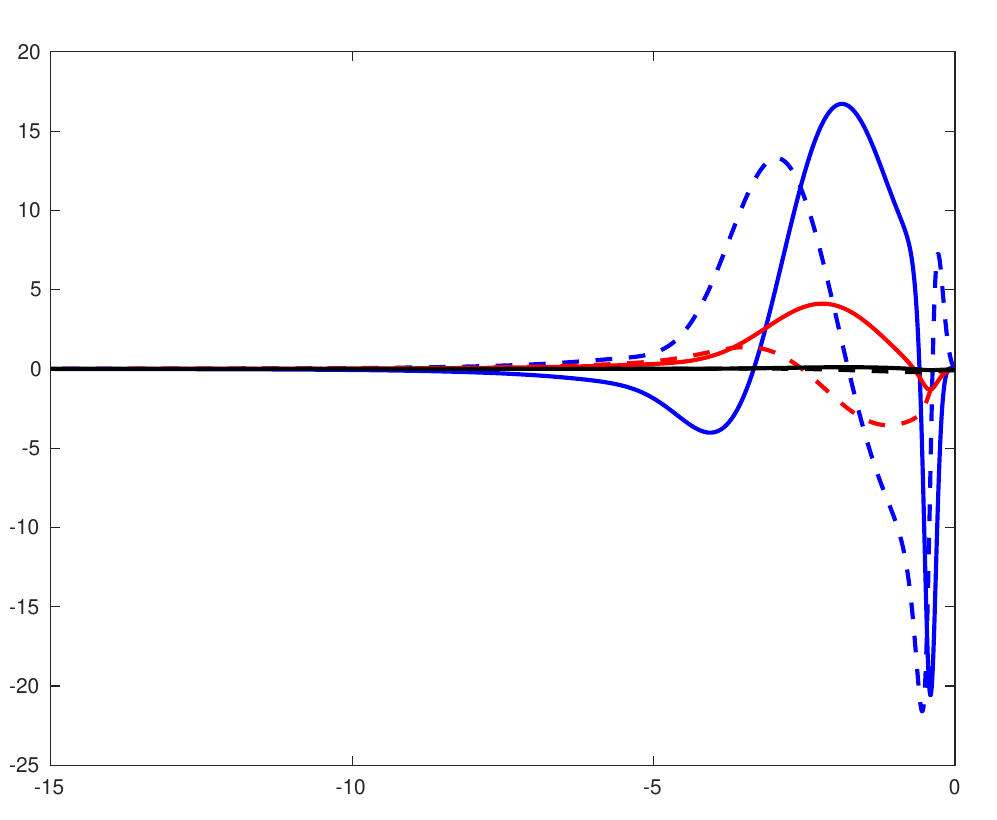}}
\caption{\label{ub_efns} 
The (asymptotic) eigenfunction on the upper branch neutral curve for $\beta=0.9$: $\hat{u}$ (red) and $\hat{v}$ (blue) on the left; $\htxx$ (blue), $\htxy$ (red) and $\htyy$ (black) on the right (in both real/imaginary parts are solid/dashed). The eigenfunction has been normalised so that $\hat{u}(0)=1$. The calculation has been done over the domain $\hY \in [-15,0]$ for clarity but larger domains (e.g. $[-50,0]$) were used for convergence purposes.} 
\end{figure}

Numerical calculations by \cite{Khalid21a} on the upper branch neutral curve suggest the scaling behaviour
\beq
(W,k,y,1-c)=\left(\frac{\hW}{\del}, \frac{\hk}{\del},\del \hY, \ha \del^2 \right)
\eeq
where all hatted variables are $O(\del^0)$ for some $\del=\del(Re) \rightarrow 0$ as $Re \rightarrow \infty$. This is the channel flow equivalent of the short-wavelength scalings for pipe flow studied by \cite{Dong22} in their \S4.1.
Rescaling the  variables (\ref{ub_mom_x})-(\ref{ub_eqn_tau_22}) as follows
\beq
(u,v,p,\tau_{11},\tau_{12},\tau_{22})
=\left( 
\hu,\hv,\frac{\hp}{\del},\frac{\htxx}{\del},\frac{\htxy}{\del},\frac{\htyy}{\del}  \right)
\eeq
leaves the problem
\begin{align}
Re \del^3 \left[ \, i \hk (\ha-\hY^2) \hu-2\hY \hv \,\right] & = -i \hk \hp+\beta (\hD^2-\hk^2) \hu + i \hk \htxx+\hD \htxy     \label{ub_mom_x}\\
Re \del^3 \left[ \, i \hk (\ha-\hY^2) \hv    \, \right] & = -\hD \hp+\beta (\hD^2-\hk^2) \hv + i \hk \htxy +\hD \htyy ,     \label{ub_mom_y}\\
i \hk \hu+\hD \hv &=0, \label{ub_incompressible} \\
\left[\frac{1}{\hW}+i \hk (\ha-\hY^2) \right] \htxx  &= 
-16 (1-\beta) \hW \hY \hv  
+16i \hk (1-\beta)\hW \hY^2 \hu \nonumber \\
& \hspace{1cm} -4(1-\beta) \hY \hD \hu
-4 \hY \htxy +\frac{2i k (1-\beta)}{\hW} \hu,               
\label{ub_eqn_tau_11}\\
\left[\frac{1}{\hW}+i \hk (\ha-\hY^2) \right] \htxy  &= 
2(1-\beta)\hv  
+8i \hk (1-\beta)\hW \hY^2 \hv 
-2\hY \htyy \nonumber \\
& \hspace{4cm}
+\frac{1-\beta}{\hW}(\hD \hu+i \hk \hv),
\label{ub_eqn_tau_12}\\
\left[\frac{1}{\hW}+i\hk(\ha-\hY^2) \right] \htyy  &= \,\,\,\,\,-4i \hk (1-\beta) \hY \hv + \frac{2 (1-\beta)}{\hW} \hD \hv,
\label{ub_eqn_tau_22}
\end{align}
where $\hD:=\partial/\partial \hY = \del D$ and no terms have been dropped. The polymer equations are invariant under this scaling (no terms are dropped) regardless of $\del$ but $\del:= Re^{-1/3}$ is forced by the momentum equation if inertia and viscous effects are to be balanced in the usual Newtonian way near a critical layer (where $\Re e(c)=U(y)$). With this choice, no terms are also dropped in the momentum equation so the scaling transformation is exact here for a parabolic base profile. The one change going from the original eigenvalue problem to this scaled version is the position of the boundary which is transformed to $\hY=\pm \infty$. Solving the asymptotic eigenvalue problem on the neutral curve is then one of finding a neutral  eigenfunction which decays away at infinity. Given the symmetry of the centre mode \citep{Khalid21a},
\beq
(u,v,p,\tau_{11},\tau_{12},\tau_{22})(-y)=(u,-v,p,\tau_{11},-\tau_{12},\tau_{22})(y),
\label{symmetry}
\eeq
it is sufficient to just impose $\hu=\hv=0$ at some large distance $\hY=-L$ ($L\gg 1$) solving over the lower half of the channel and imposing appropriate symmetry across $y=0$ ($L=15$ to $50$ were used to explore convergence at $\beta=0.9$): see eigenfunctions in figure \ref{ub_efns}.  

The asymptotic properties $(\hW,\hk,\,\Re e(\ha))$ of the upper branch neutral curve  are given by seeking
\beq
\max_{\hk} \,\{\,\hW \,| \,\min_{\hk}\Im m[\ha(\hk,\hW)] =0 \,\}
\label{ub_problem}
\eeq
in the eigenvalue problem (\ref{ub_mom_x})-(\ref{ub_eqn_tau_22}), that is, by finding the largest value of $\hW$ for which there are no unstable eigenfunctions (the growth rate $kc_i=-\Im m(\ha)/Re$) ). The required $\hk$ and $\ha$ are defined by the neutral eigenfunction at this maximum. The results of this procedure for $\beta=0.9$ are that  
\beq
W \sim 9.1725Re^{1/3}, \quad k \sim 0.5023 Re^{1/3}, \quad c_r \sim 1-0.0908Re^{-2/3}
\label{upperscalings}
\eeq
on the upper branch neutral curve as $Re \rightarrow \infty$. Table 1 and figure \ref{ub_lb_curves} show that this asymptotic result is good down to at least $Re=100$. Using the elasticity number $E:=W/Re$ as in \cite{Khalid21a},
these scalings are equivalent to $Re \sim O(E^{-3/2}$), $k \sim O(E^{-1/2})$ and $1-c \sim O(E)$ as $E \rightarrow 0$ which is consistent with figure 11 in \cite{Khalid21a}.

%
%
\begin{table}
\begin{center}
\begin{tabular}{@{}rrccccc@{}}
   $Re$   \hspace{0.25cm} & \hspace{1cm} &                $\hat{W}$ & \hspace{0.2cm}   & $\hat{k}$  &  \hspace{0.2cm}    &  $\hat{a}$   \\ \hline
                   &&             &&              &&          \\
    100            &&    8.9409   &&  0.4912      &&  0.1018 \\
   200             &&    9.1809   &&  0.5010      &&  0.0908 \\ 
   1,000           &&    9.1730   &&  0.5020      &&  0.0909 \\
 10,000            &&    9.1727   &&  0.5022      &&  0.0908\\
                   &&             &&              &&              \\
$\infty$           &&    9.1725   &&  0.5023      &&  0.0908   \\[6pt]
\end{tabular}
\end{center}
%

\caption{\label{Table1} Upper branch neutral curve characteristics for $\beta=0.9$ as $Re \rightarrow \infty$. At a given finite $Re$, $\hat{W}=W/Re^{1/3}$, $\hat{k}=k/Re^{1/3}$ and $\hat{a}=(1-c)Re^{2/3}$. The bottom line shows the values given in (\ref{upperscalings}) found by solving the asymptotic problem in (\ref{ub_problem}).}
\end{table}

%
%

%
%
\subsection{Lower branch in $Re$ vs $W$ plane at fixed $\beta$ \label{lower}}

\begin{figure}
\centering
\scalebox{0.38}[0.38]{\includegraphics{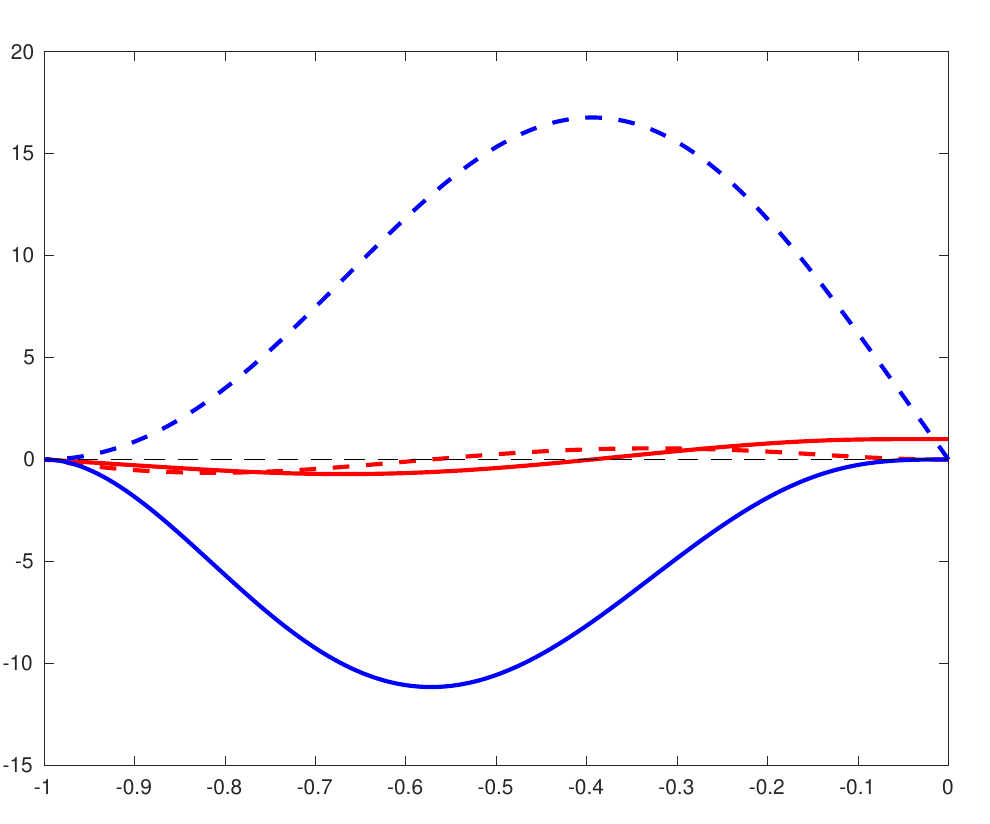}}
\scalebox{0.38}[0.38]{\includegraphics{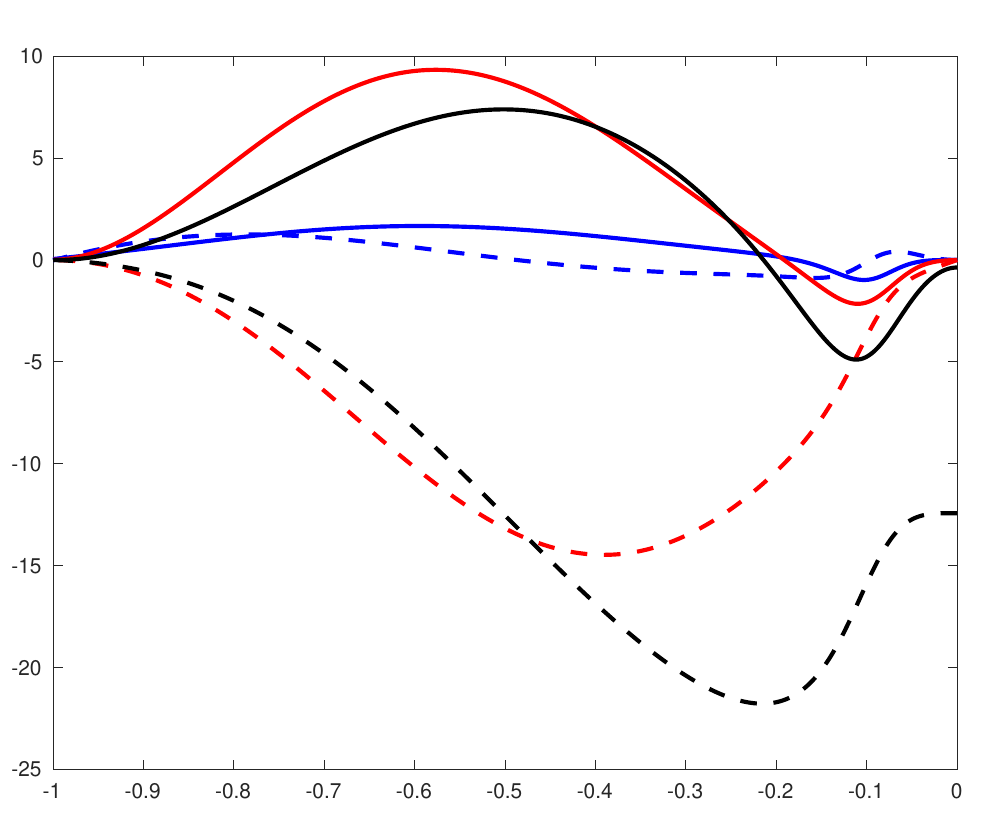}}\\
\scalebox{0.38}[0.38]{\includegraphics{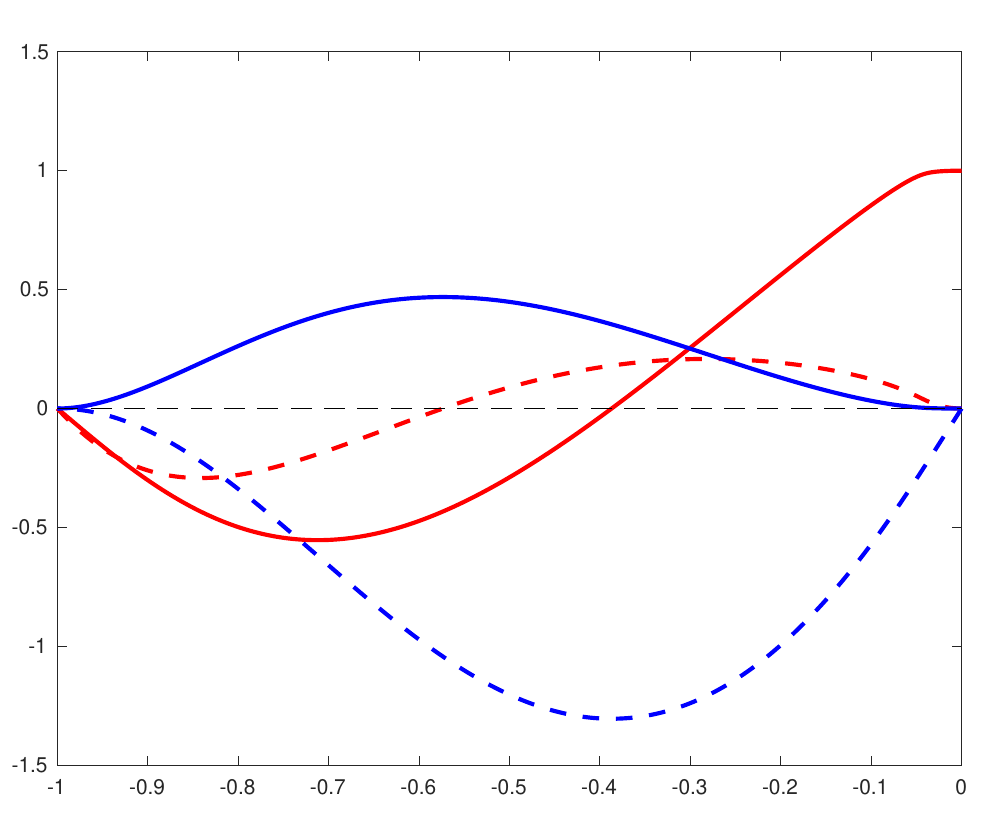}}
\scalebox{0.38}[0.38]{\includegraphics{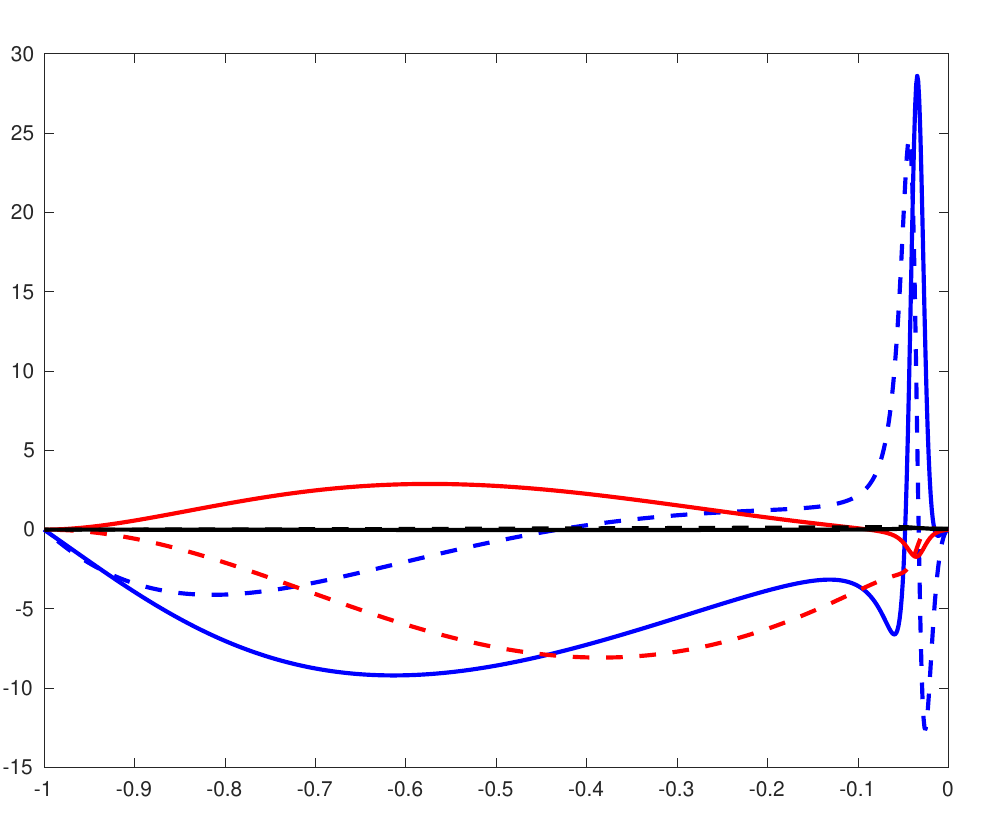}}\\
\caption{\label{lb_efns} 
The (asymptotic) eigenfunctions on the lower branch neutral curve for $\beta=0.9$ (top row) and $\beta=0.994$ (bottom row): $\hat{u}$ (red) and $\hat{v}$ (blue) on the left; $\htxx$ (blue), $\htxy$ (red) and $\htyy$ (black) on the right (in both real/imaginary parts are solid/dashed). Both eigenfunctions have been normalised so that $\hat{u}(0)=1$.} 
\end{figure}

Numerical calculations on the lower  branch neutral curve (\cite{Khalid21a}) suggest a long-wavelength limit scaling of the following form
\beq
(u,v,p,\tau_{11},\tau_{12},\tau_{22}, W, k)
=\left( 
\hu,\frac{\hv}{Re},
Re\, \hp_0+\hp_1+O(Re^{-1}),
Re\,\htxx,\htxy,\frac{\htyy}{Re}, Re\, \hW, \frac{\hk}{Re}
\right)
\eeq
where all hatted variables are $O(1)$ as $Re \rightarrow \infty$, $\hp_0$ is a constant ($D\hp_0=0$) and $c$ stays $O(1)$ and bounded away from $0$ (the wall advection speed) and $1$ (the centreline advection speed): see figure \ref{ub_lb_curves} (bottom left). In their long-wavelength analysis for pipe flow (their \S3), \cite{Dong22} only consider $1/Re \ll k \ll 1$ and so do not treat the lower branch neutral curve where again $k=O(1/Re)$ is found numerically (\cite{Garg18}).
With these rescalings, (\ref{ub_mom_x})-(\ref{ub_eqn_tau_22}) become
\begin{align}
i \hk (1-y^2-c)\hu-2y\hv & = -i \hk \hp_0+\beta D^2 \hu + i \hk \htxx+D \htxy     \label{lb_mom_x}\\
i \hk (1-y^2-c)\hv \hspace{1cm} & =-D\hp_1+\beta D^2 \hv + i \hk \htxy+D \htyy ,      \label{lb_mom_y}\\
i \hk \hu+ D\hv &=0, \label{lb_incompressible}
\end{align}
for the velocity field and, for the polymer stress,
\begin{align}
\left[\frac{1}{\hW}+i \hk (1-y^2-c) \right] \htxx  &= -16 (1-\beta) \hW y \,\hv +16i \hk (1-\beta) \hW y^2\,\hu -4y\, \htxy\nonumber \\
& \hspace{5.5cm}-4(1-\beta)y D \hu ,               
\label{lb_eqn_tau_11}\\
\left[\frac{1}{\hW}+i \hk (1-y^2-c) \right] \htxy  &= \,2(1-\beta) \hv +8i \hk (1-\beta)\hW y^2\,\hv -2y \, \htyy+\frac{1-\beta}{\hW}D\hu,
\label{lb_eqn_tau_12}\\
\left[\frac{1}{\hW}+i \hk (1-y^2-c) \right] \htyy  &= \,\,\,\,\,-4i \hk (1-\beta) y \,\hv + \frac{2 (1-\beta)}{\hW} D\hv.
\label{lb_eqn_tau_22}
\end{align}
Since $D\hp_0=0$, differentiating (\ref{lb_mom_x}) leads directly to the vorticity equation
\beq
\beta D^4 \hv= -\hk^2 D\htxx+i\hk D^2\htxy +i \hk \left[ (1-y^2-c)D^2 \hv+2 \hv\right]     \label{lb_vorticity}
\eeq
and (\ref{lb_mom_y}), which just defines $\hp_1$, can be ignored.  The problem defined by (\ref{lb_incompressible})-(\ref{lb_vorticity}) is then an eigenvalue problem for $c$.  Since there is no rescaling of the spatial dimension, the neutral eigenfunction is global and easily resolved. The asymptotic properties $(\hW,\hk,\,c$ of the lower branch neutral curve  are given by seeking
\beq
\max_{\hk} \,\{\,\hW \,| \,\max_{\hk}\Im m[c(\hk,\hW)] =0 \,\}
\label{lb_problem}
\eeq
and the results are shown in Table 2. The asymptotic scalings  for $\beta=0.9$ where $Re \rightarrow \infty$ are the same as for $\beta=0.994$ where $Re \rightarrow -\infty$ but the eigenfunctions looks distinctly different in the polymer stress field - see figure \ref{lb_efns}. 

The lower branch scalings are apparent in figure 13 of \cite{Khalid21a} (see also their figure 18).  The upper branch asymptote is reached for $Re_c \rightarrow \infty$ and $E \rightarrow 0$ whereas the vertical asymptote $E \rightarrow E_\infty$ (a finite value) as $Re_c \rightarrow \infty$ corresponds to the lower branch asymptote and the results in Table 2 can be used to predict $E_c$. For  example $(1-\beta)E_c=(1-\beta)\hW = 0.1058$ at $\beta=0.9$ and $(1-\beta)E_c=(1-\beta)\hW = 0.352$ at $\beta=0.98$ (note $E_c \lesssim \max_{Re_c}{E}$ for a given $\beta$ in figure 13 in \cite{Khalid21a}). \cite{Khalid21b} (their figure 4) show that there is no asymptote for $\beta > 0.990552$. Instead the asymptote has to flip to $Re \rightarrow -\infty$ and $E_c < 0$ as shown for example with $\beta=0.994$ in figure \ref{ub_lb_curves} (bottom right).

%
%
\begin{table}
\begin{center}
\begin{tabular}{@{}lrccrcc@{}}
 \hspace{0.1cm}  $\beta$   \hspace{0.25cm} & \hspace{1cm} &                $\hat{W}$ & \hspace{0.2cm}   & $\hat{k}$\hspace{0.25cm}  &  \hspace{0.2cm}    &  $c$   \\ \hline
           &&             &&              &&          \\
   0.9     &&  1.058    &&  62.17     &&  0.999553  \\ 
   0.98    &&  17.60\quad \quad    &&  13.32     &&  0.998788\\
   0.994   &&  -128.1 \quad \quad \quad  &&   -5.97    &&  0.999198\\
           &&             &&              &&      \\
\end{tabular}
\end{center}
%

\caption{\label{Table2}Lower branch neutral curve characteristics for $\beta=0.9$ and $0.98$ as $Re \rightarrow \infty$ and for $\beta=0.994$ as $Re \rightarrow -\infty$ (hence $\hW$ and $\hk$ are both negative). A comparison with results from finite $Re$ calculations are shown in figure \ref{ub_lb_curves} (bottom right). }
\end{table}

%
%

%
%

\section{$W \rightarrow \infty$ Asymptotics for Inertialess ($Re=0$) Channel Flow}

Without inertia ($Re=0$), the relevant asymptotic limit is $W \rightarrow \infty$ and $\beta \rightarrow 1$ such that $W(1-\beta)=\lam$ is an $O(1)$ constant, e.g. see insets A and  B of figure 2 in  \cite{Khalid21b} (or figure 8 in their supplementary material) which suggest $3.5 \lesssim \lam \lesssim 10$ for instability. Physically, of course, this means $W$ is large but finite whereas $Re$ can be considered separately  as small as desired but is strictly not zero as there is  flow. The point is mathematically, the $Re \rightarrow 0$ limit is regular so it is convenient to set $Re=0$ to get the true limiting values of key dependencies (e.g. how $1-c$ scales with $W$ on the neutral curve).

\begin{figure}
\centering
\scalebox{0.55}[0.55]{\includegraphics{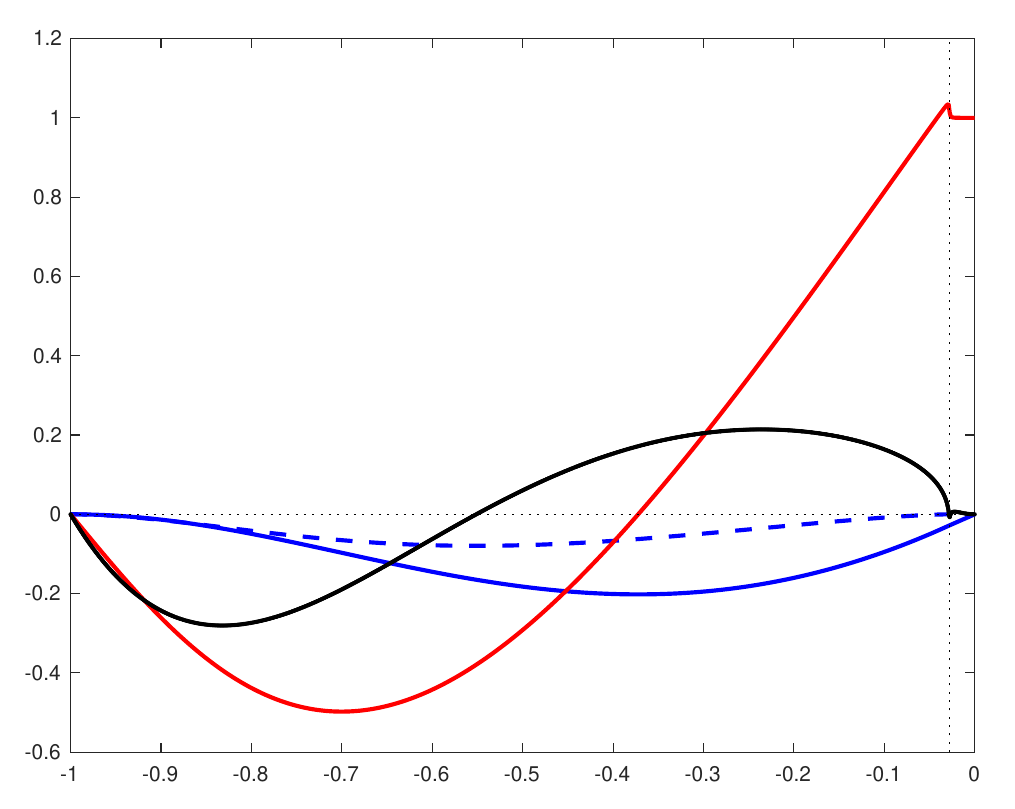}}
\scalebox{0.55}[0.55]{\includegraphics{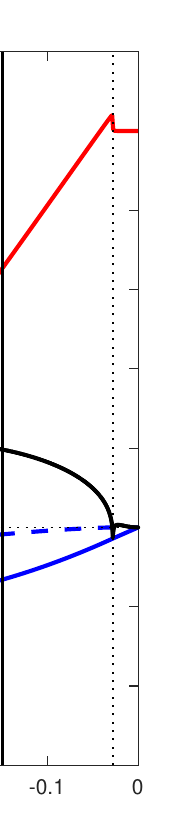}}
\scalebox{0.55}[0.55]{\includegraphics{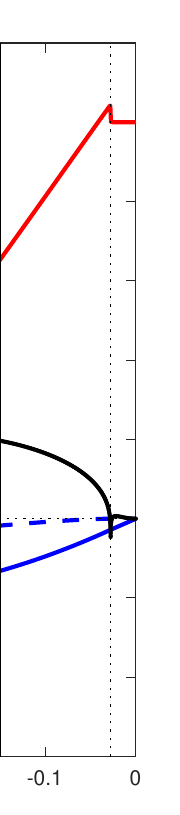}}
\caption{\label{fig1} Neutral eigenfunction at W=32,000 (left full plot), 128,000 (middle partial plot) and 512,000 (right partial plot), $k=1.1$ and $\lam \approx 4.11=W(1-\beta)$ and $c_r \approx 0.99218$. Vertical dotted line near $y=0$ is the critical layer where $U=1-y^2=c_r$.  $v$ is blue (real/imaginary parts solid/dashed respectively),  real part of $Dv$ is red and imaginary part is black.  Notice the O(1)  jump in $\Re e(Dv)$ (and in $\Re e (D^2v)$)  across the critical layer and  the increasingly singular behaviour of $\Im m(Dv)$ (black solid line) as $W$ increases. }
\label{eigenfunction}
\end{figure}

As already mentioned in \S 3 and $Re >0$, the centre mode instability has a certain symmetry about the midplane $y=0$: $u$ is symmetric and $v$ antisymmetric: see (\ref{symmetry}). Henceforth we only consider $y \in [-1,0]$ and impose non-slip boundary conditions $v(-1)=Dv(-1)=0$ at the solid lower plate and symmetry conditions $v(0)=D^2v(0)=0$ at the midplane.  Numerically (see Appendix A for details), we find on the neutral curve that the eigenfunction has a critical layer near the midplane across which $v$ is continuous but there are jumps in $\Re e(Dv)$ and $\Re e(D^2v)$ and singular-looking behaviour for $\Im m(Dv)$ where the phase of the eigenfunction is set by making $\Re e(Dv)=1$ at the midplane: see figure \ref{eigenfunction}.

A key issue is whether the critical layer at $y=y^*$, defined by $U(y^*)=c_r$ so $y^*:=-\sqrt{1-c_r} \in[-1,0]$,  approaches the midplane as it does in classical Orr-Sommerfeld analysis for Newtonian shear flows \citep{DrazinReid} or not. Certainly figure \ref{eigenfunction} suggests `not' and earlier (pre-shooting code) attempts to develop the asymptotic structure could not reconcile $c_r \rightarrow 1$ as $W \rightarrow \infty$ with an O(1) jump in $D^2v$ across the critical layer. This means a  novel aspect of the asymptotics here  is that despite $1-c_r$ being very small, it does in fact remain O(1) as $W \rightarrow \infty$: see Table 3.

%
%
\begin{table}
\begin{center}
\begin{tabular}{@{}lrrccccc@{}}
  & $W$   \hspace{0.25cm}      & \hspace{0.5cm}  &                $c_r$     & $k\Lambda_{-}$                       &\hspace*{1cm}     & $c_r$  &  $k\Lambda_{+}$   \\ \hline
    &                &                             &                                 &                                                   &                             &             &                                               \\
 $k=0.1$ &   32,000    & &          0.998390     &      3.41331  &&      0.999893   &  7.62705  \\
 &  128,000    &&           0.998408     &     3.39349    &&      0.999943   &  9.25320  \\
 & 512,000    &&          0.998412     &      3.38849    &&      0.999955   &  9.76344 \\
 & *2,048,000   &&         0.998413     &       3.38724    &&      0.999957   &  9.80010 \\
& *8,192,000  &&         0.998413      &      3.38693     &&     0.999957   &   9.81447 \\
& *32,768,000 &&        0.998413      &        3.38686    &&   0.999957   &   9.81826   \\
&            &&                       &                  &&              &             \\
& $\infty$ \hspace{0.25cm}    && 0.998413 &   3.38683     &&   0.999957    &   9.81954    \\
 &                            &&          &               &&              &             \\
$k=1.1$ &    64,000   &&    0.9992186       &   4.526249            &&   0.999853  &    7.63580       \\
  & 128,000    &&    0.9992183       &   4.523676            &&   0.999855  &    7.65132       \\
  & 256,000    &&    0.9992181       &   4.522393            &&   0.999855  &    7.65932     \\
  & *1,024,000   &&    0.9992180       &   4.521432            &&   0.999856  &    7.66538       \\
  & *8,192,000   &&    0.9992180       &   4.521154            &&   0.999856  &    7.66715         \\
  &                  & &       &   &&      &                \\
& $\infty$ \hspace{0.25cm}  &&   0.9992180 &    4.521114     &&   0.999856  &    7.66740          \\[6pt]
\end{tabular}
\end{center}
%

\caption{\label{Table3} $c_r$ and $k\Lambda:=k(1-\beta)W$ as plotted in inset A of figure 2 in  \cite{Khalid21b} on the neutral curve for  $k=0.1$  and $1.1$ at various large $W$ ($\pm$ indicates upper and lower parts of the neutral curve and * results computed using quadruple rather than double precision).   The $\infty$ entry comes from Richardson extrapolation eliminating the leading $O(1/W)$ error. In all cases $\Lambda$ is slower to converge than $c_r$ with the upper curve calculations on the right highlighting this. }
\end{table}

We introduce a small parameter
\beq
\eps :=\frac{1}{W}
\label{eps}
\eeq
and take the distinquished limit $\beta=1-\eps \Lambda$ where $\Lambda$ is an O(1) number to be determined.
The eigenfunction plotted in figure \ref{eigenfunction} shows abrupt changes in the solution as it crosses a critical layer at $y=c_r$. The solution either side of the critical layer - the `outer' solution - must satisfy the governing equations with $\beta=1$, $\eps=0$ and $\lam=O(1)$ with the critical layer supplying appropriate `matching' conditions between the two parts.  The purpose of the asymptotic analysis developed below is  to identify  analytic expressions for these matching conditions so that the two parts of the outer solution can be fitted together in the limit of $W \rightarrow \infty$ without solving for the critical layer. This defines the leading solution to the problem which includes the leading $O(W^0)$ value of $c$.

%
%
%
%
\subsection{Outer solution}

We refer to the  `outer' solution as the solution in the regions $y-y^*=O(1)$.  Assuming $v=O(1)$, then $\tau_{11}$ and $\tau_{12}$ are $O(1)$ whereas $\tau_{22}=O(1/W)$. The leading order outer problem is then 
\begin{align}
(D^2-k^2)^2 v &= -k^2 D \tau_{11}+ik(D^2+k^2)\tau_{12},\\
iku+Dv            &=0,\\
ik(U-c) \tau_{11}  &= -v DT_{11} +2ikT_{11}u+2U' \tau_{12},     \label{outer_t11}\\
ik(U-c) \tau_{12}  &= ik T_{11}v,                                                   \label{outer_t12}        
\end{align}
which can be simplified to the 4th order problem
\beq
(D^2-k^2)^2 v=-ikD\left[  T_{11} D \left( \frac{v}{U-c} \right)\right]+ \frac{ik^3 T_{11}v}{U-c}.
\label{interior}
\eeq
For $y < y^*$,  outer boundary conditions are  $v(-1)=0, Dv(-1)=0$  and for $y> y^*$ $v(0)=0=D^2v(0)$ with the critical layer supplying 4 matching conditions (for $v$, $Dv$, $D^2v$ and $D^3v$ respectively). This will produce a well posed  eigenvalue problem for $c(\lam,k)$. Ultimately, the problem is to find $\min \lam $ over all pairs $(\lam,k)$ where $c_i(\lam,k)=0$.

%
%
\subsection{Inner solution}

The inner solution  is  the solution in the critical layer  $y-y^*=O(\eps)$ where the thickness comes from balancing polymer advection and relaxation processes: see the left hand sides of the perturbed polymer stress equations (\ref{eqn_tau_11})-(\ref{eqn_tau_22}). We therefore define a critical layer variable $Y$ and the corresponding derivative $\hD$,
\beq
Y:=\frac{y-y^*}{\eps}, \qquad
D=\frac{1}{\eps} \hD:= \frac{1}{\eps} \frac{\partial}{\partial Y},
\eeq
so that, for example, $1/W+ik(U-c)=\eps(1+ik\Us Y+ik U_*^{''}\eps Y^2)$ where $\Us:=U^{'}(y^*)=-2y^*>0.$
Then the appropriate expansions turn out to be
\begin{align}
v     &=              \hspace{0.5cm}     \hv_0 +\eps \log \eps \,\Pi Y +\eps \,\hv_1(Y)+ \eps^2\log  \eps \biggl( \hv_\ell(Y)+\tfrac{1}{2} \Om Y^2 \biggr)+\eps^2 \,\hv_2(Y) \nonumber \\
& \hspace{5cm}+ \eps^3 \log \eps \biggl( \hv_{2\ell}+ \tfrac{1}{6} \Upsilon Y^3 \biggr)+ \eps^3 \,\hv_3(Y)+\ldots
\label{v_crit}\\
u    &= \frac{i}{k} \left[ \hspace{1.1cm} \log \eps \,\Pi  + \hD \hv_1(Y) +\eps \log \eps \biggl( \hD \hv_\ell(Y)+\Om Y \biggr) + \eps \hD \hv_2(Y) \right. \nonumber \\ 
& \hspace{4.5cm} + \eps^2 \log \eps \biggl( \hD \hv_{2 \ell} + \half \Upsilon Y^2 \biggr) +\eps^2 \hD \hv_3(Y)+ \ldots   \
\label{u_crit}\\
Du & = \frac{i}{k} \left[ \hspace{2.3cm} \frac{1}{\eps} \hD^2 \hv_1(Y)+\log \eps \biggl( \hD^2\hv_\ell(Y)+\Om \biggr)+\hD^2 \hv_2(Y) \right. \nonumber \\
& \hspace{5cm}+\eps \log \eps \biggl( \hD^2 \hv_{2 \ell}+  \Upsilon Y\biggr)+\eps \hD^2 \hv_3(Y)+\ldots  \label{Du_crit}  \\
D^2u & = \frac{i}{k} \left[ \hspace{2.3cm} \frac{1}{\eps^2} \hD^3 \hv_1(Y)+\frac{\log \eps}{\eps} \hD^3\hv_\ell(Y)+\frac{1}{\eps}\hD^3 \hv_2(Y) \right. \nonumber \\
& \hspace{5.5cm} +\log \eps  \biggl( \hD^3 \hv_{2 \ell}+ \Upsilon \biggr) +\hD^3 \hv_3(Y)+\ldots   \label{D2u_crit}
\end{align}
for the velocity fields and 
\begin{align}
\tau_{11} &= \frac{1}{\eps^2} \htau_{11}^0(Y)+ \frac{\log \eps}{\eps} \htau_{11}^\ell(Y)+ \frac{1}{\eps} \htau_{11}^1(Y)+\log \eps \, \htau_{11}^{2 \ell}+ \htau_{11}^2(Y)+\ldots, \label{tau11}\\
\tau_{12} &= \frac{1}{\eps} \htau_{12}^0(Y)+ \log \eps \,\htau_{12}^\ell(Y)+ \htau_{12}^1(Y)+\eps \log \eps\, \htau_{12}^{2 \ell}+\eps \htau_{12}^2(Y)+\ldots, \label{tau12}\\
\tau_{22} & = \htau_{22}^0(Y)+\eps \log \eps \,\htau_{22}^\ell(Y) +\eps \htau_{22}^1(Y)+\eps^2 \log \eps \, \htau_{22}^{2\ell}+\eps^2 \htau_{22}^2(Y)+\ldots \label{tau22}
\end{align}
for the polymer stresses where $\Pi$, $\Om$  and $\Upsilon$ are complex constants which will emerge below. The reason we need to go so deep into these expansions is the outer problem is 4th order and therefore  requires  jump conditions down to $D^3v$ {\em plus} there is a singularity at the critical layer. Together these require considering the equation for $\hD^3 \hv_3(Y)$ which is $O(\eps^2)$ down  in the expansion of $D^3v$ i.e. we need to go to third order in the expansion. The intermediate $O(\log \eps)$ terms are formally needed to fix up the logarithmic terms which arise but turn out to be unimportant for deriving the matching conditions.

To keep track of the influence of  $U'$ and $U^{''}$ when we probe later why plane Couette flow doesn't have a neutral curve,  it is useful to expand the base state  around $y=y_*$ in the critical layer as follows,
\begin{align}
U          &= U_*+\eps Y  \Us+ \half \eps^2 Y^2 U_*^{''}+\ldots, \label{U}\\
DU        &= \Us+\eps Y U_*^{''}+\ldots, \label{DU}\\
T_{11}   &= T_{11}^{*(0)}+ \eps Y T_{11}^{*(1)}+\eps^2 Y^2 T_{11}^{*(2)}+ \ldots,  \label{T11}\\
DT_{11} &=  T_{11}^{*(1)}+2\eps Y T_{11}^{*(2)} + \dots, \label{DT11}\\
T_{12}   &= \hspace{1cm}      + \eps T_{12}^{*(1)} \hspace{0.25cm}+ \eps^2 Y T_{12}^{*(2)}+\ldots, \label{T12}\\
DT_{12} &= \hspace {1cm}    +\eps T_{12}^{*(2)}+ \ldots \label{DT12}\\
\end{align}
where 
\beq
T_{11}^{*(0)}:= 2 \lam  U_*^{'2}, \quad T_{11}^{*(1)}:= 4\lam \Us U_*^{''}, \quad T_{11}^{*(2)}:= 2 \lam U_*^{''2}
\eeq
and
\beq
T_{12}^{*(1)}:= \lam \Us, \quad T_{12}^{*(2)}:= \lam U_*^{''}
\eeq
assuming that $U_*^{'''}=0$ for simplicity (true for both channel and Couette flow).

Now we substitute expansions (\ref{v_crit})-(\ref{D2u_crit}) for the perturbation velocity field, (\ref{tau11})-(\ref{tau22}) for the perturbation polymer stresses and (\ref{U})-(\ref{T12}) for the base state into the equations (\ref{incompressible})-(\ref{vorticity}) and collect similar order terms to create a hierarchy of problems in the usual way.

%
%
\subsection{Leading order: $O(1/\eps^3)$ in the  Stokes equation}

At leading order,
\begin{align}
X \htau_{22}^0  & = 2ik T_{12}^{*(1)} \hv_0                               \label{tau_22^0}\\
X\htau_{12}^0  & = ik T_{11}^{*(0)} \hv_0+\Us \htau_{22}^0              \label{tau_12^0}\\
X\htau_{11}^0  & = 2\Us \htau_{12}^0                                \label{tau_11^0}\\
\hD^4\hv_1 & = -k^2 \hD \htau_{11}^0+ik \hD^2 \htau_{12}^0
\end{align}
where $X:=(1+ik\Us Y)$.
Integrating the last equation twice with respect to $Y$ gives
\beq
D^2v=\frac{1}{\eps}\hD^2 \hv_1= 
-\frac{k^2}{\eps} \int \htau_{11}^0 dY+ \frac{ik}{\eps} \htau_{12}^0 = 
\frac{k^2 T_{11}^{*(0)} \hv_0}{\eps X} \sim 
\frac{-ik T_{11}^{*(0)} \hv_0}{\Us(y-y^*)}  \quad {\rm as}\quad Y\rightarrow \pm \infty
\label{D2v}
\eeq
as the $O(1/\eps)$ integration constants must be zero otherwise the solution cannot be matched with the central region.  This leading solution for $D^2v$ immediately suggests that the asymptotic matching will be a challenge. There is a simple pole singularity in $D^2v$ at the critical layer and as a consequence, a double pole singularity in $D^3v$. Since a double pole is symmetric across the  critical layer, it does not enter into the matching conditions for $D^3v$ but will certainly obscure any matching criterion present involving higher order, less singular behaviour.

Forewarned, we press on and integrate once to  give
\beq
Dv= \hD \hv_1= \frac{-ik T_{11}^{*(0)} \hv_0}{\Us}\biggl[ \log X +\alpha_1 \biggr]
\label{Dv_lead}
\eeq
where $\alpha_1$ is another complex constant). Matching to the exterior requires that a \\
$-ikT_{11}^{*(0)}\hv_0/\Us \log \eps$ term is present in the expression for $Dv$ so then
\beq
Dv \sim \frac{-ikT_{11}^{*(0)} \hv_0}{\Us} \biggl[  \log [ ik\Us (y-y^*)] +\alpha_1  \biggr] \quad {\rm for} \quad \eps \ll |y-y^*| 
\eeq
This means that the first complex coefficient $\Pi$ in the expansions (\ref{v_crit})-(\ref{D2u_crit}) has to be $\Pi=-ikT_{11}^{*(0)} \hv_0/\Us=-2ik\lam \Us \hv_0$ and so
\beq
Dv \rightarrow \left\{\begin{array}{cc}  \Pi  \biggl( \log |k\Us (y-y^*)|  +\alpha_1-\tfrac{1}{2}i \pi  \, \biggr)  & y  \rightarrow y^{*-}\\
&                                  \\
\Pi \biggl( \log |ik\Us(y-y^*)|   +\alpha_1+\tfrac{1}{2} i\pi   \, \biggr)  & y \rightarrow y^{*+}                \label{Dv1}
\end{array} \right.
\eeq
This gives a jump in $Dv$ across the critical layer of $i \pi \Pi$.  Integrating  (\ref{Dv_lead}) gives
\beq
\hv_1=\Pi \left[   \frac{X \log X}{ik\Us} +(\alpha_1-1)Y\right]
\eeq
since $\hv_1(y_*)=0$ as $v(y_*)=\hv_0$ by definition.
%

%
%
\subsection{Next order: $O(\log \eps/\eps^2)$ in the  Stokes equation} 

Working to next order
\begin{align}
\htau_{22}^\ell                     & = \frac{2ik T_{12}^{*(1)} \Pi Y+2\lam \Pi}{X}=2  \lam \Pi\\
\htau_{12}^\ell  & = \frac{ik T_{11}^{*(0)} \Pi Y+\Us \htau_{22}^\ell }{X}= 2 \lam \Pi \Us \\
\htau_{11}^\ell  & = \frac{2 \Pi (-T_{11}^{*(0)}X+2ik U_*^{'2} T_{12}^{*(1)}+2 U_*^{'2} \lam ) }{X^3}=0
\end{align}
so $\hD^4\hv_\ell = 0$ and therefore $\hv_\ell=0$.

%
%
\subsection{$O(1/\eps^2)$ in the  Stokes equation}

This is the order which will complete the jump condition in $D^2v$. At $O(1/\eps^2)$, we get
\begin{align}
X \htau_{22}^1  & =-\half ik U_*^{''} Y^2 \htau_{22}^0
+2ik T_{12}^{*(2)} \hv_0
+2ik T_{12}^{*(1)} \hv_1
+2 \lam \hD \hv_1
\label{a22^1}\\
X  \htau_{12}^1  & = -\half ik U_*^{''} Y^2 \htau_{12}^0
+ ( ik T_{11}^{*(1)}Y-T_{12}^{*(2)} )\hv_0
+ik T_{11}^{*(0)}\hv_1
+\Us \htau_{22}^1
+U_*^{''} Y \htau_{22}^0
+\tfrac{i \lam}{k} \hD^2 \hv_1
\label{a12^1}\\
X \htau_{11}^1  & = 
-\half ik U_*^{''} Y^2 \htau_{11}^0
-T_{11}^{*(1)} \hv_0
-2 T_{11}^{*(0)} \hD \hv_1
+\tfrac{2i}{k} T_{12}^{*(1)} \hD^2 \hv_1
+2 \Us \htau_{12}^{1}+2 U_*^{''} Y\htau_{12}^0
\label{a11^1}
\end{align}
with
\beq
\hD^4\hv_2 = \lam \hD^4 \hv_1-k^2 \hD \htau_{11}^1+ik \hD^2 \htau_{12}^1.
\eeq
Integrating twice
\beq
\hD^2\hv_2 = \lam \hD^2 \hv_1-k^2 \int \htau_{11}^1\, dY +ik \htau_{12}^1+\Theta Y+\Phi.
\label{D2v}
\eeq
where $\Theta$ and $\Phi$ are complex constants. $\Theta Y$ is unmatchable in the interior so $\Theta$ must be $0$. In terms of deriving jump conditions across the critical layer, the presence of the constant $\Phi$ means we are  only interested in the asymptotic behaviour (as $Y \rightarrow \pm \infty$) of the RHS of (\ref{D2v}) which gives rise to jumps across the layer. With this in mind, it is straightforward to show from (\ref{a22^1}) and (\ref{a12^1}) that
\beq
ik \htau_{12}^1 = \frac{k^2 [T_{11}^{*(0)}]^2}{U_*^{'2}} \hv_0 \log X+const \qquad {\rm as} \quad Y \rightarrow \pm \infty
\label{a12_contribution}
\eeq
The other term on the RHS,
\begin{align}
-k^2 \int \htau_{11}^1 \,dY = \int & 
\overbrace{   \frac{\half  k^2 U_*^{''} Y^2 \htau_{11}^0}{\Us X}   }^{(i)}
\overbrace{  -\frac{i k T_{11}^{*(0)} \hv_0}{\Us X}  }^{(ii)}
\overbrace{-\frac{ 2ikT_{11}^{*(0)}  \hD \hv_1}{\Us X}}^{(iii)}  \nonumber \\
& 
-\frac{2 T_{12}^{*(1)} \hD^2 \hv_1}{\Us X}
+\underbrace{\frac{2ik U_*^{''} Y \htau_{12}^0}{\Us X}}_{(iv)}
+\underbrace{\frac{2 ik \htau_{12}^1}{X}}_{(v)},
\end{align}
is more involved with each labelled term contributing.  Respectively, as $Y \rightarrow \pm \infty$,
\begin{align}
(i)  & \rightarrow  - \frac{ik U_*^{''} T_{11}^{*(0)}}{U_*^{'2}}\hv_0 \log X      \nonumber \\
(ii)  & \rightarrow  -\frac{k T_{11}^{*(1)}}{\Us} \hv_0 \log X \nonumber \\
(iii)  & \rightarrow   - \frac{2k^2 [T_{11}^{*(0)}]^2\,\hv_0}{U_*^{'2}} \biggl[ \half (\log X)^2+\alpha_1 \log X \biggr] \nonumber \\
(iv)  & \rightarrow   \frac{2ik U_*^{''} T_{11}^{*(0)}\hv_0}{U_*^{'2}} \log  X \nonumber 
\end{align}
The $(v)^{th}$ term has to be further subdivided as follows
\begin{align}
\frac{ik}{\Us} \int \frac{2\Us \htau_{12}^1}{X} \,dY 
=\int & 
\overbrace{\frac{k^2 U_*^{''} Y^2 \htau_{12}^0}{X^2}}^{a}
-\frac{ 2ik \lam U_*^{''} \hv_0}{X^2}
+\overbrace{\frac{-2k^2 T_{11}^{*(1)}  Y \hv_0}{X^2}}^{b}  \nonumber \\
& 
+\underbrace{\frac{-2k^2 T_{11}^{*(0)} \hv_1}{X^2}}_{c}
+\frac{2ik \Us \htau_{22}^1}{X^2}
+\frac{ 2ik U_*^{''} Y\htau_{22}^0}{X^2}
-\frac{2\lam \hD^2 \hv_1}{X^2},
\end{align}
with the respective asymptotic behaviour
s $Y \rightarrow \pm \infty$,
\begin{align}
(v)_a  & \rightarrow   -\frac{ik U_*^{''} T_{11}^{*(0)} \hv_0}{U_*^{'2}} \log X     \nonumber \\
(v)_b  & \rightarrow    \frac{2ik T_{11}^{*(1)} \hv_0}{\Us} \log X \nonumber \\
(v)_c  &  \rightarrow  \frac{2k^2 [T_{11}^{*(0)}]^2\,\hv_0}{U_*^{'2}} \biggl[ \half (\log X)^2+(\alpha_1-1) \log X \biggr] \nonumber 
\end{align}
Adding all the contributions
\begin{align}
\hD^2 \hv_2  &\sim 
\biggl[ 
\frac{-k^2 [T_{11}^{*(0)}]^2}{U_*^{'2}}+\frac{ik T_{11}^{*(1)}}{\Us}
\biggr] \log X +const \nonumber \\
&\hspace{3cm}= (\,4ik \lam U_*^{''} -4k^2 \lam ^2 U_*^{'2}\,) \hv_0 \log [ik\Us(y-y_*)]+const, \nonumber\\
&\hspace{3cm}=\Om \log [ik\Us(y-y_*)]+const. \label{d2v2}
\end{align}
Therefore the second complex coefficient $\Om$ in the expansions (\ref{v_crit})-(\ref{D2u_crit}) has to be $\Om=(\,4ik \lam U_*^{''} -4k^2 \lam ^2 U_*^{'2}\,) \hv_0$ and this indicates the finite jump in $D^2v$ across the critical layer. Integrating  (\ref{d2v2}) twice gives
\begin{align}
\hD \hv_2 & =  -\frac{i\Om}{k \Us} \biggl( X \log X-X\biggr)+ \alpha_2, \nonumber \\
      \hv_2  & = -\frac{\Om}{k^2 U_*^{'2}} \biggl( \frac{1}{2}X^2 \log X-\frac{3}{4}X^2\biggr)+\alpha_2Y+\beta_2,
\end{align}
where $\alpha_2$ and $\beta_2$ are constants. Since $\hv_2=0$ at $Y=0 \,(X=1)$ by definition of $\hv_0$, $\beta_2=-3\Om/(4k^2 U_*^{'2})$.

%
%
\subsection{$O(\log \eps/\eps)$ in the  Stokes equation}

The $O(\log \eps/\eps)$ Stokes equation balance integrated once gives
\beq
\hD^3 \hv_{2 \ell}= -k^2 \htau_{11}^{2 \ell}+i k \hD \htau_{12}^{2 \ell}+ const
\label{v3}
\eeq
as $\hv_{\ell}=0$. Now, since
\begin{align}
\htau_{22}^{2 \ell} &=
\frac{1}{X} \biggl(  
-\half i k U_*^{''} Y^2 \htau_{22}^\ell + ik \Om T_{12}^{*(1)} Y^2+2ik T_{12}^{*(2)} Y^2 \Pi+2 \lam \Om Y \biggr) 
\sim \frac{Y^2}{X}, \nonumber\\
\htau_{12}^{2 \ell} &=
\frac{1}{X} \biggl(  
-\half ik U_*^{''} Y^2 \htau_{12}^\ell +(ikT_{11}^{*(1)}Y-T_{12}^{*(2)})\Pi Y+\half ik \Om T_{11}^{*(0)}Y^2\nonumber \\
&\hspace{6cm} +U_*^{''} Y \htau_{22}^{\ell}+\Us \htau_{22}^{2 \ell}+\frac{i \lam \Om}{k}
 \biggr) \sim \frac{Y^2}{X}, \nonumber\\
\htau_{11}^{2 \ell} & =
\frac{1}{X} \biggl(  
-3 \Pi Y T_{11}^{*(1)}-2T_{11}^{*(0)} \Om Y+\frac{ 2i T_{12}^{*(1)}\Om}{k}+2U_*^{''} Y \htau_{12}^\ell+2\Us \htau_{12}^{2 \ell}
 \biggr) \sim O\biggl(\frac{Y^2}{X^2}, \frac{Y}{X} \biggr) \nonumber
\end{align}
as $|Y| \rightarrow \infty$, the RHS of (\ref{v3}) generates at best constant terms as $Y \rightarrow \pm \infty$ which can be removed by the `const'. Hence there is no consequence outside the critical layer as expected (the $\log \eps$ ordered terms are there just to fix up the logarithmic terms). Note, however, $\hv_{2\ell}$ is non trivial in the critical layer but it is just doesn't drive a jump across it.

%
%
\subsection{$O(1/\eps)$ in the  Stokes equation}

This is the order at which a finite jump in $D^3 v$ across the critical layer is determined. The $O(1/\eps)$ Stokes equation balance integrated once gives
\beq
\hD^3 \hv_3=\underbrace{2k^2\hD\hv_1}_{(a)}+\lam \hD^3 \hv_2 -\underbrace{k^2 \htau_{11}^2}_{(d)} +k^2\htau_{22}^0
+\overbrace{i k \hD \htau_{12}^2}^{(c)}+\overbrace{i k^3 \int \htau_{12}^0 dY}^{(b)}
\label{eqn_v3}
\eeq
Since we are only interested in jumps across the critical layer, we focus on the logarithmic terms appearing on the RHS of
(\ref{eqn_v3}) (only the labelled terms contribute). Terms (a) and (b) follow immediately
\begin{align}
(a)  & \sim -\frac{2 i k^3T_{11}^{*(0)} \hv_0}{\Us} \log X            \quad ({\rm from} \,\, (\ref{Dv_lead})  )   \nonumber \\    
(b)  & \sim \frac{ i k^3T_{11}^{*(0)} \hv_0}{\Us} \log X \quad ({\rm from} \,\, (\ref{tau_12^0})  )    \label{D3v_1}
\end{align}
as $|Y| \rightarrow \infty$.
Terms (c) and (d) require more work going to yet higher order in the polymer stress equations (\ref{eqn_tau_11})-(\ref{eqn_tau_22}).
%
%
Starting with (c),
\begin{align}
ik\hD \htau_{12}^2 &= 
\overbrace{  \hD \biggl[\frac{\half k^2 U_*^{''} Y^2 \htau_{12}^{*(1)} }{ X } \biggr]}^{(i)} 
+ \overbrace{ \hD \biggl[\frac{ ( -k^2 T_{11}^{*(1)}Y-ikT_{12}^{*(2)} )\hv_1 }{ X }\biggr]  }^{(ii)}
+\overbrace{  \hD \biggl[-\frac{k^2 T_{11}^{*(0)} \hv_2}{ X } \biggr] }^{(iii)}
\biggr. \nonumber\\
& \hspace{1.5cm} 
+\hD \biggl[ 
\frac{ik \Us \htau_{22}^2}{X}+\frac{ -k^2( \lam+ Y^2 T_{11}^{*(2)} ) \hv_0}{ X } 
+\frac{ik U_*^{''}Y \htau_{22}^{1}}{X}-\frac{ \lam \hD^2 \hv_2}{X}
\biggr]
\end{align}
Then the labelled terms have the following logarithmic behaviour
\begin{align}
(i)   & \sim -\frac{ k^2 U_*^{''} [T_{11}^{*(0)}]^2 \hv_0}{2U_*^{'3}} \log X             \nonumber \\    
(ii)  & \sim   \frac{ k^2 T_{11}^{*(0)} T_{11}^{*(1)} \hv_0 }{ U_*^{'2} } \log X                 \nonumber\\
(iii) & \sim   \frac{ ik T_{11}^{*(0)}  \Om }{2 \Us} \log X    \label{D3v_2}
\end{align}
as $|Y| \rightarrow \infty$.
Now considering (d),
\begin{align}
-k^2 \htau_{11}^2 &=\frac{\half ik^3 U_*^{''}Y^2 \htau_{11}^1}{X}
+\frac{2k^2 T_{11}^{*(2)}Y \hv_0}{X}
+\overbrace{ \frac{k^2 T_{11}^{*(1)} \hv_1}{X} }^{(a)}
+\overbrace{ \frac{k^2 T_{11}^{*(0)} \hD \hv_2}{X} }^{(b)} 
+\overbrace{ \frac{k^2 T_{11}^{*(1)} Y\hD \hv_1}{X} }^{(c)} 
 \nonumber \\
& \hspace{0.5cm}
-2\frac{ik^2T_{12}^{*(1)} \hD^2 \hv_2}{X}
-2\frac{ik^2T_{12}^{*(2)} Y\hD^2 \hv_1}{X}
+\underbrace{ \frac{-2k^2 \Us \htau_{12}^2  }{X } }_{(d)}
+\underbrace{ \frac{-2k^2 U_*^{''} Y \htau_{12}^1}{X} }_{(e)}
\end{align}
with labelled terms having the following logarithmic behaviour
\begin{align}
(a)  & \sim   -\frac{k^2 T_{11}^{*(0)} T_{11}^{*(1)}  \hv_0}{U_*^{`2}} \log X             \nonumber \\    
(b)  & \sim   -\frac{2 i k T_{11}^{*(0)} \Om }{\Us } \log X                 \nonumber\\
(c)  & \sim   -\frac{ 2 k^2T_{11}^{*(0)} T_{11}^{*(1)} \hv_0}{U_*^{'2}} \log X \nonumber\\
(d)  & \sim   \biggl[ 
-\frac{k^2 U_*^{''} [T_{11}^{*(0)}]^2 \hv_0}{U_*^{'3}} 
+\frac{2k^2 T_{11}^{*(0)}T_{11}^{*(1)} \hv_0}{U_*^{'2}}
+\frac{ik T_{11}^{*(0)} \Om}{\Us}    
\biggr] \log X  \nonumber \\
(e) & \sim  \frac{2k^2 U_*^{''} [T_{11}^{*(0)}]^2 \hv_0}{U_*^{'3}} \log X   \label{D3v_3}
\end{align}
as $|Y| \rightarrow \infty$. Bringing the expressions from (\ref{D3v_1}), (\ref{D3v_2}) \& (\ref{D3v_3}) together,
\beq
\hD^3 \hv_3 \sim \Upsilon \log X:=
\biggl[  
k \Pi(k -2i \lam)-ik\lam \Us \Om    
\biggr] \log X  \quad {\rm as} \, \, Y \rightarrow \pm \infty
\label{v3_final}
\eeq
which gives a finite jump in $D^3 v$ across the critical layer.
Integrating repeatedly gives
\begin{align}
\hD^2 \hv_3 &= -\frac{i\Upsilon}{k \Us} \biggl(X \log X -X \biggr)+\alpha_3, \nonumber\\
\hD     \hv_3 &= -\frac{\Upsilon}{k^2 U_*^{'2}}\biggl( \frac{1}{2} X^2 \log X-\frac{3}{4} X^2\biggr)+\alpha_3 Y +\beta_3,\nonumber\\
           \hv_3 &= \frac{i \Upsilon}{k^3 U_*^{'3}}\biggl( \frac{1}{6} X^3 \log X-\frac{11}{36}X^3\biggr)+\frac{1}{2}\alpha_3 Y^2 +\beta_2 Y+\gamma_3.  \label{v3}
\end{align}
where $\alpha_3, \beta_3$ and $\gamma_3$ are constants with $\gamma_3$ set by ensuring that $\hv_3=0$ at $Y=0\, (X=1)$).

%
%
\begin{figure}
\centering
\scalebox{0.7}[0.7]{\includegraphics{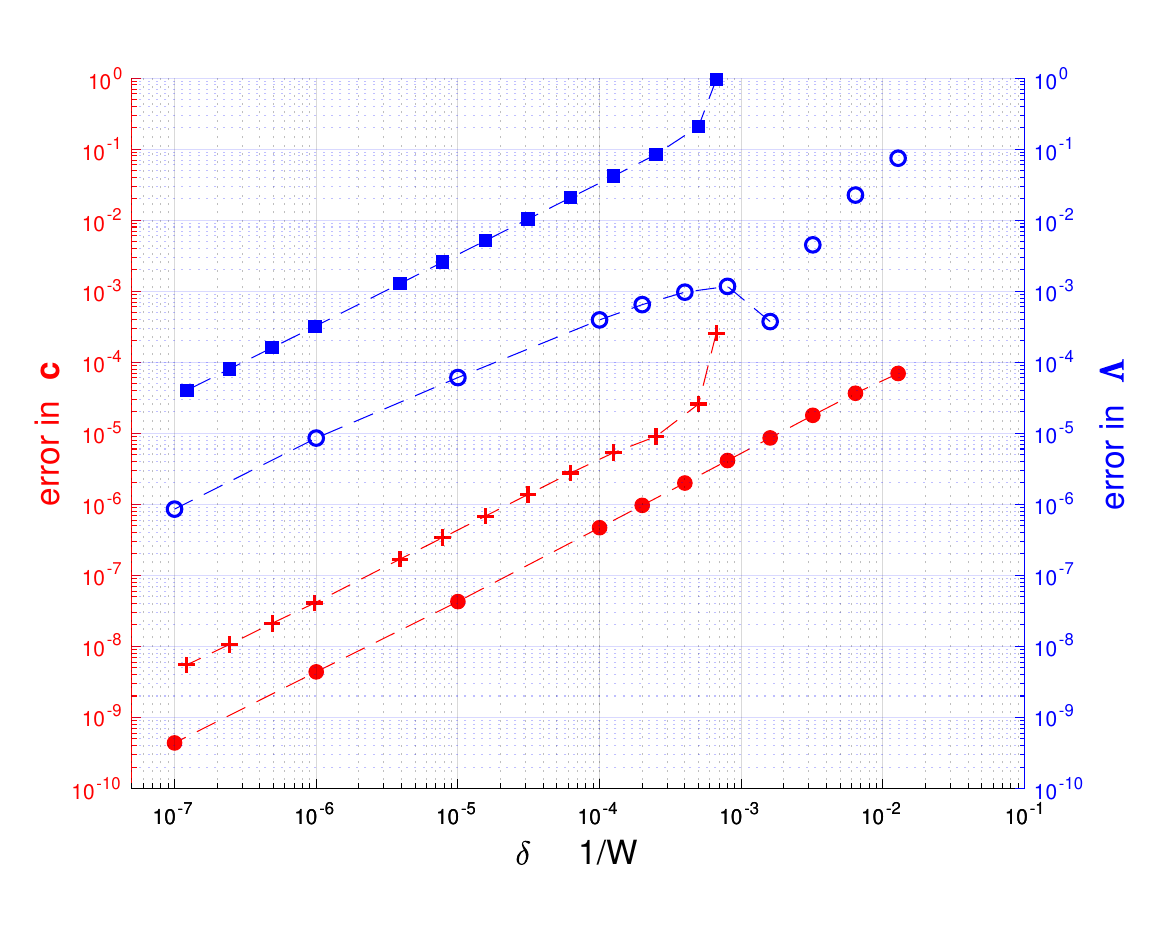}}
\caption{\label{error} The `error' in $c$ and $\Lambda$ as a function of finite $\delta$ compared to their $\delta=0$ limiting values for the asymptotically matched solution and for the numerical solution as a function of $1/W$ compared to the $W\rightarrow \infty$ limit for $k=1.1$ and $(c, \lam)=(0.999218, 4.1101)$.  Circles are used for the  asymptotically matched solutions (filled red circles for $c$ and open blue circles for $\lam$); for the numerical approximations,  red $+$ is used for $c$ and blue filled squares for $\lam$.  The `truth' $(c(0),\Lambda(0))$ is  estimated by assuming $c(0)=c(\del) +a\del+b \del^2+\ldots$ and eliminating the leading error between the two most accurate predictions - i.e.  $c(0):= (\,10c(10^{-7})-c(10^{-6})\,)/9$ (Note $1/W \leq 1/974$ for an Oldroyd-B fluid  (\cite{Khalid21b})).}
\end{figure}

%
%
\begin{figure}
\centering
\scalebox{0.55}[0.55]{\includegraphics{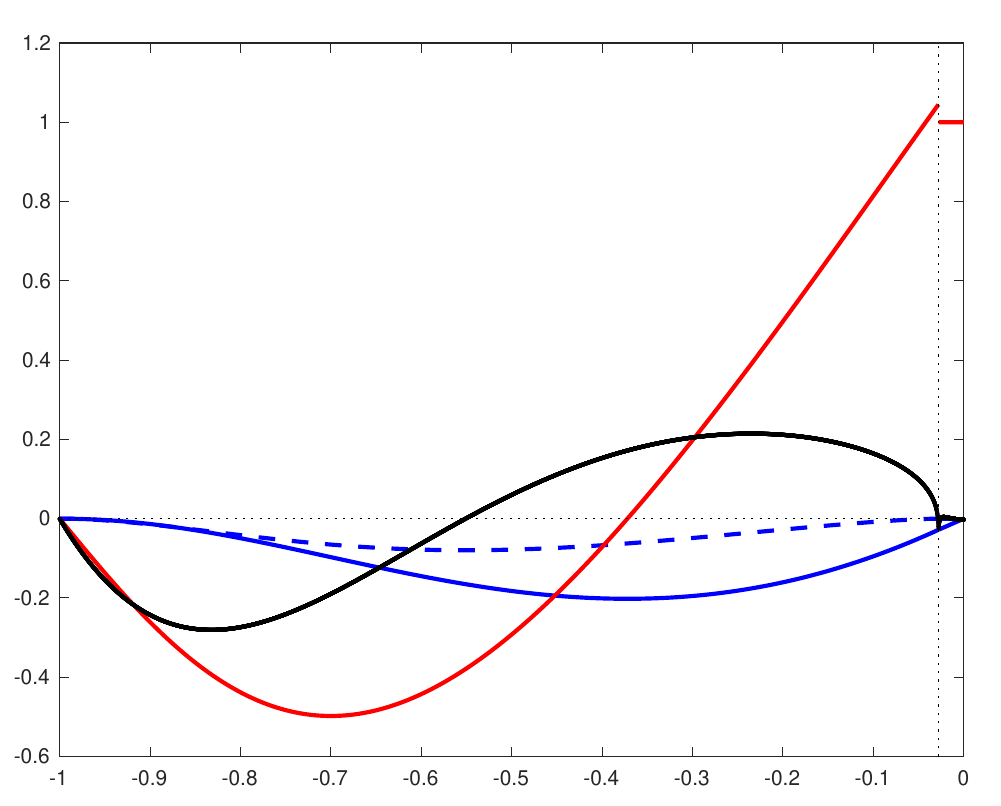}}
\scalebox{0.55}[0.55]{\includegraphics{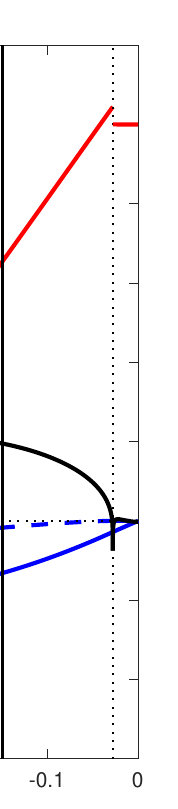}}
\caption{\label{asymptotic}  The outer solution at $k=1.1$ and $\lam \approx 4.11=W(1-\beta)$ found by applying the matching conditions (\ref{D3v_jump})-(\ref{Dv+}) with $\del=2 \times 10^{-4}$ (full plot) and $\del=5 \times 10^{-6}$ (partial right plot).  As in figure 1, $v$ is blue (real/imaginary parts solid/dashed respectively),  real part of $Dv$ is red and imaginary part is black.  Notice the increasingly singular behaviour in the imaginary part of $Dv$ at the critical point as $\del$ decreases.}
\end{figure}

%
%
\begin{figure}
\centering
\scalebox{0.37}[0.37]{\includegraphics{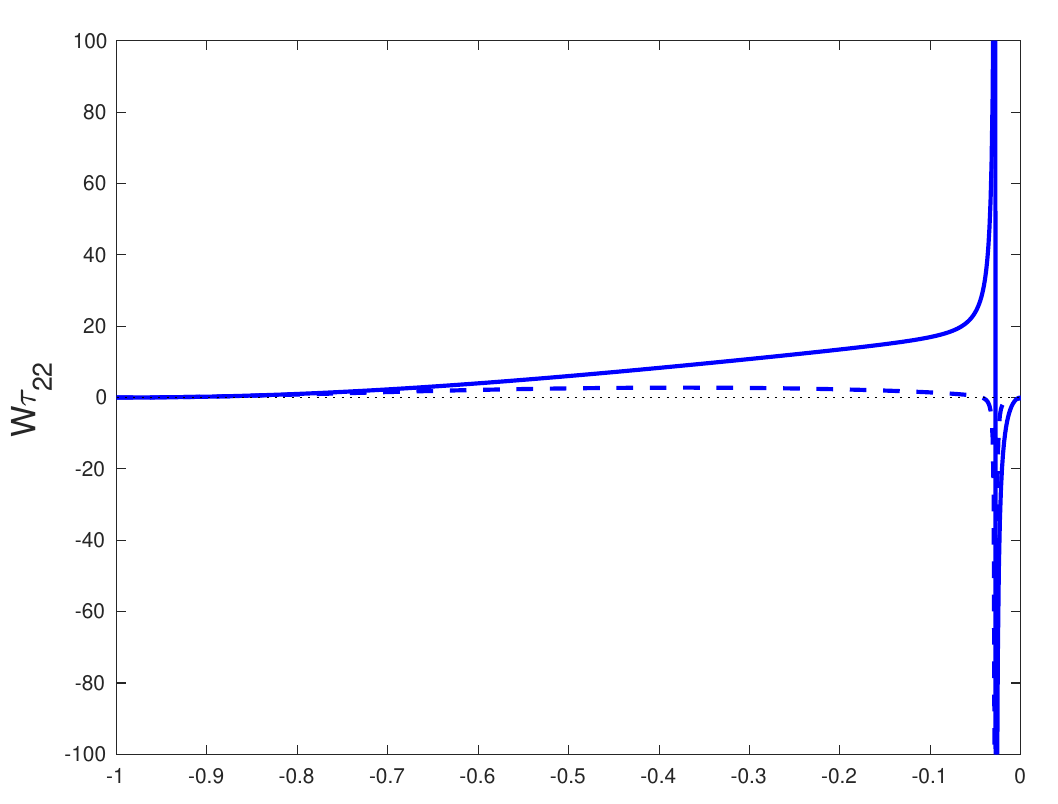}}
\scalebox{0.37}[0.37]{\includegraphics{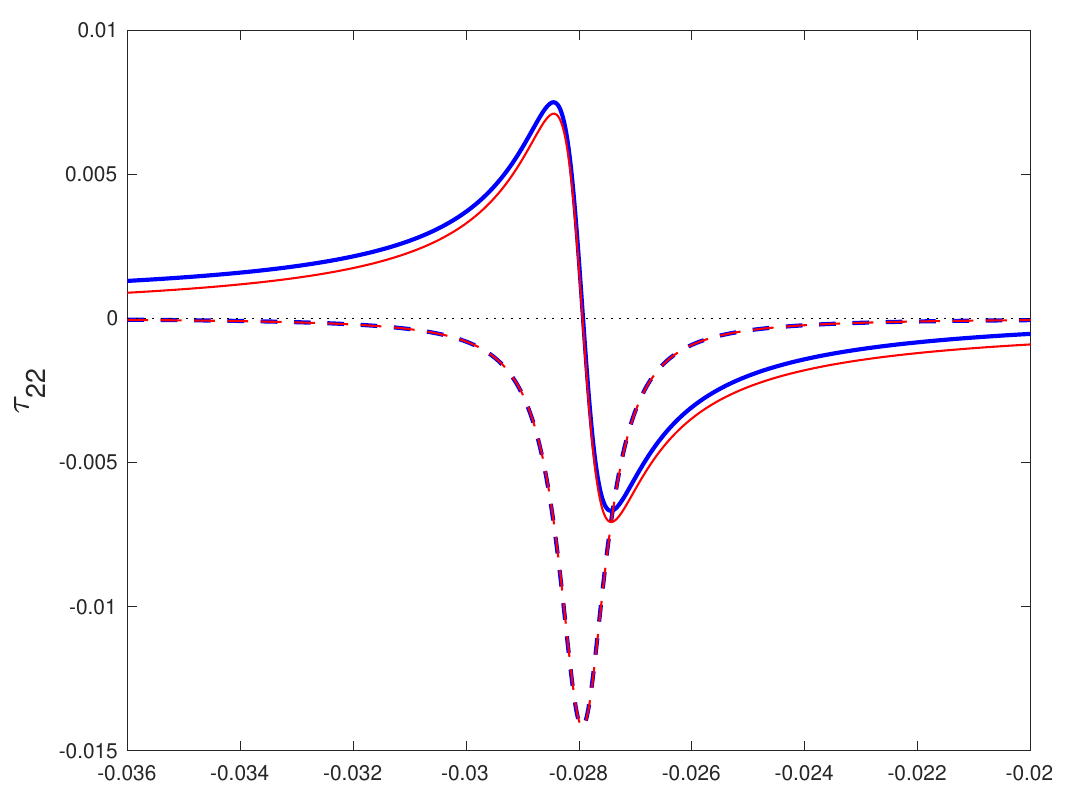}}\\
\scalebox{0.37}[0.37]{\includegraphics{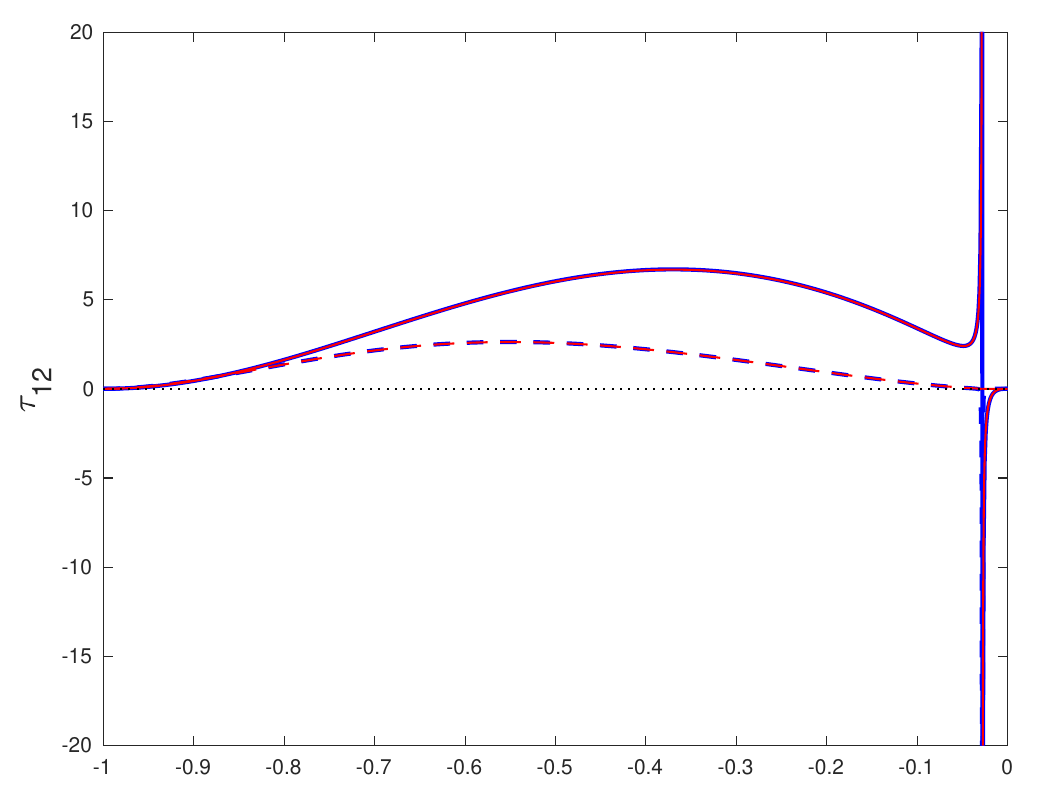}}
\scalebox{0.37}[0.37]{\includegraphics{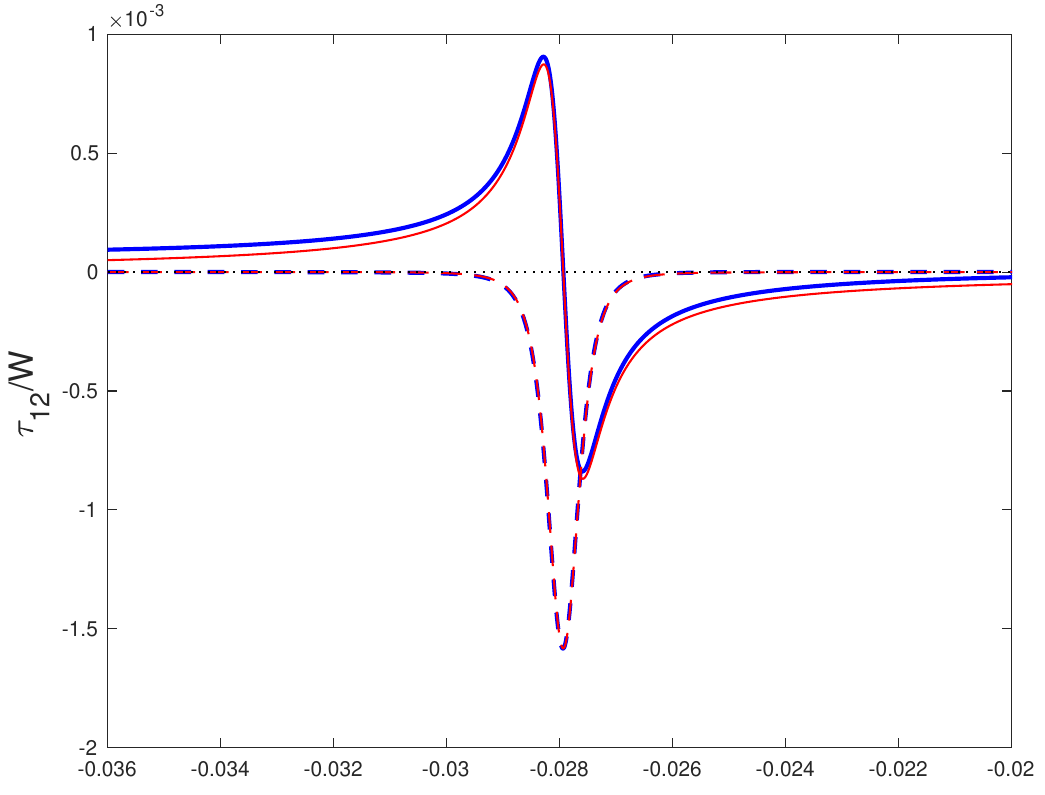}}\\
\scalebox{0.37}[0.37]{\includegraphics{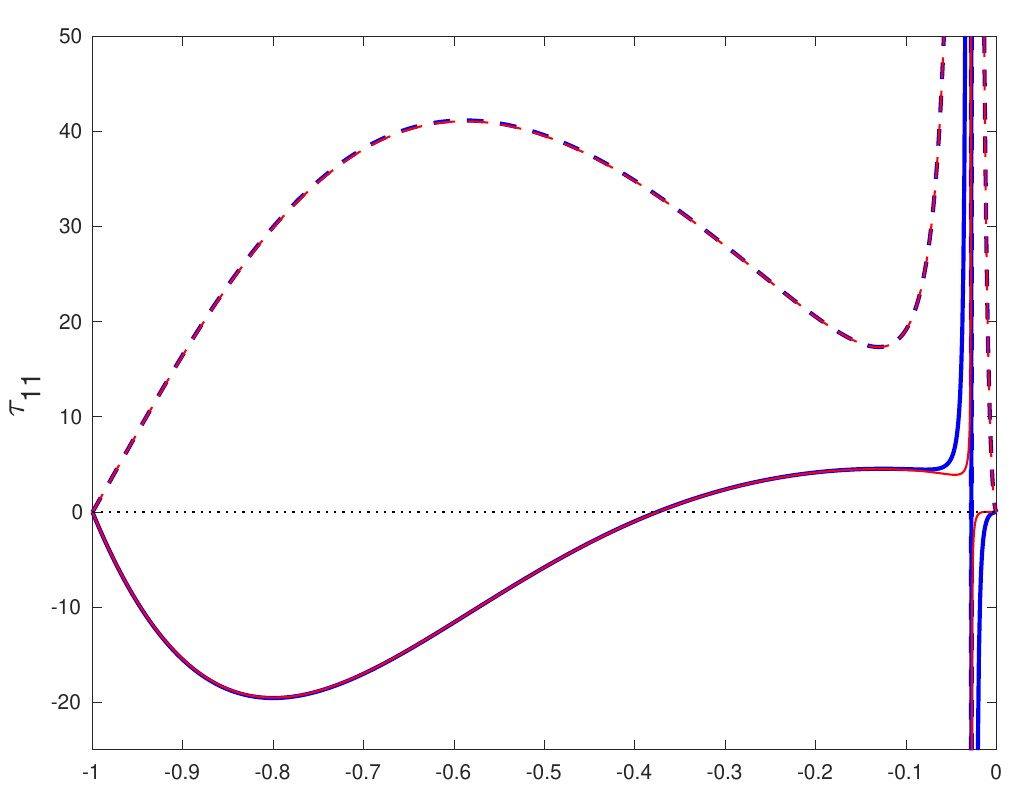}}
\scalebox{0.37}[0.37]{\includegraphics{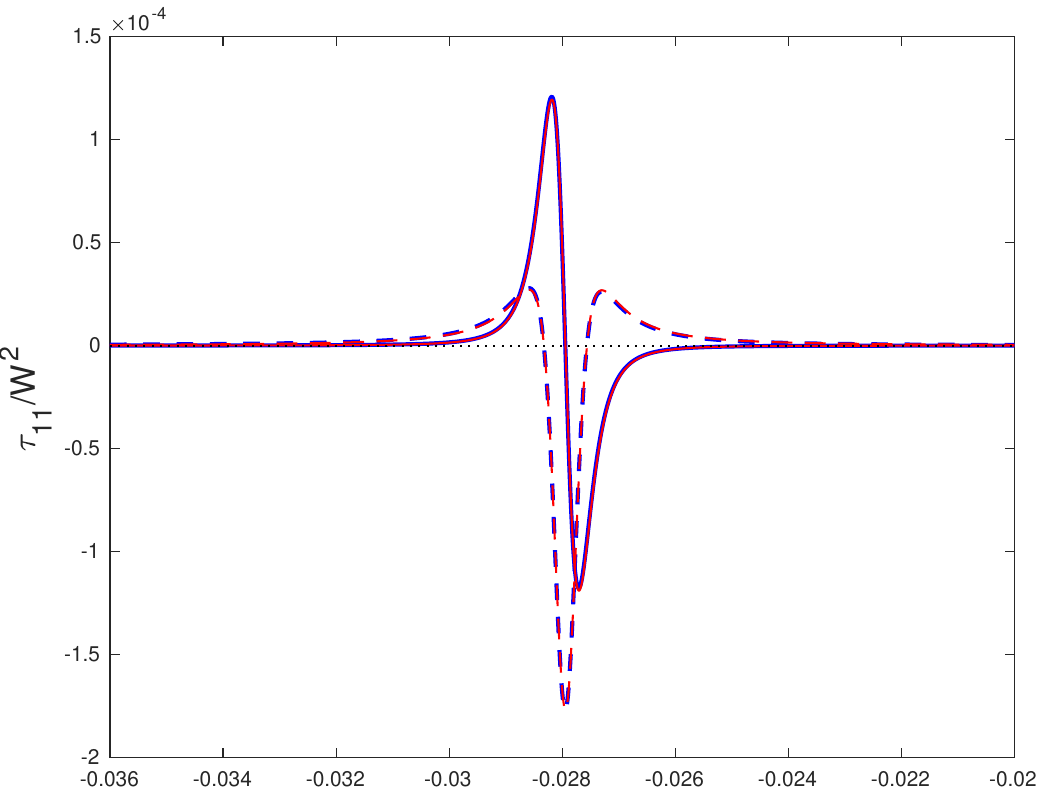}}\\
\caption{\label{stress} A comparison of the full stress field  at $k=1.1$, $\lam \approx 4.11$ computed numerically at $W=32,000$ (in blue both left and right columns) with the outer solutions (left column) and leading inner asymptotic solutions given by (\ref{tau_22^0})-(\ref{tau_11^0}) (right column - note the rescaling of both axes),  both marked in red.  Solid lines indicate real parts and dashed lines imaginary parts.) Note there is no $\tau_{22}$ component in the leading order outer solution hence the absence of a red line in the top left plot.}
\end{figure}

%
%
\subsection{Matching conditions across the critical layer}

All the work above has built up a high order (in $\eps$) representation for $\hv_0$  as follows
\begin{align}
v & =  \hv_0 +\eps \,\biggl [\hv_1(Y)+ \Pi Y \log \eps \biggr]
+\eps^2 \,\biggl[ \hv_2(Y)+ \tfrac{1}{2} \Om Y^2\ \log \eps \biggr] + \eps^3 \log \eps \,\hv_{2\ell}\nonumber \\
& \hspace{7cm}+ \eps^3 \biggl[ \,\hv_3(Y)+\tfrac{1}{6} \Upsilon Y^3 \,\biggr] +\ldots.
\end{align}
What's important is the behaviour as $y \rightarrow y^*$ for matching to the outer solutions; specifically
\begin{align}
v &=\hv_0+\biggl[ \Pi (y-y^*)   +\frac{1}{2} \Om (y-y^*)^2+ \frac{1}{6}\Upsilon (y-y^*)^3+\ldots \biggr] \log[ ik \Us(y-y^*)] \nonumber \\
& \hspace{2cm} +\Pi (y-y^*)(\alpha_1-1)-\frac{3}{4} \Om (y-y^*)^2-\frac{11}{36} \Upsilon (y-y^*)^3+O(\eps,(y-y^*)^4)\label{v}
\end{align}
where all the constants disappear at leading order in $\eps$ with the exception of $\alpha_1$.
From this,  the various  jump conditions
\beq
[A]^{+}_{-}:=A(y^*+\del)-A(y^*-\del), 
\eeq
where $\del \rightarrow 0$, can be deduced as 
\begin{align}
[v]^{+}_{-}  &=  2\Pi \del [ \,\log|k\Us \del|+\alpha_1-1\,]+\tfrac{1}{2} i \pi  \Om \del^2 +\Upsilon \del^3 [\, \tfrac{1}{3}\log |k \Us \del |-\tfrac{11}{18} \,],  \label{v_jump}\\
[Dv]^{+}_{-} &=i \pi \Pi +2\del \Om [ \,\log |k \Us \del|-1 \,]+\half i \pi \Upsilon \del^2 \label{Dv_jump}\\
[D^2v]^{+}_{-} &=\frac{2 \Pi}{\del}+i \pi \Om +2 \del \Upsilon [  \,\log |k \Us \del| -1\, ], \label{D2v_jump}\\
[D^3v]^{+}_{-} &=\frac{2\Om}{\del}+ i \pi \Upsilon .    \label{D3v_jump}
\end{align}
with
\beq
\Pi       = -2 ik \lam \Us \hv_0  ,\quad
\Om       =  (\,4ik \lam U_*^{''} -4k^2 \lam ^2 U_*^{'2}\,) \hv_0 ,\quad
\Upsilon  =  k \Pi(k -2i \lam)-ik\lam \Us \Om.
\label{constants}
\eeq
The conditions (\ref{v_jump})-(\ref{D3v_jump}) are correct to 3 orders in $\del$. This is clear in all but the last jump condition. Here $D^3v \sim -\Pi/\del^2$ which, since it is even in $\del$, does not appear in the required jump. 

The interior problem  (\ref{interior}) is 4th order with 2 boundary conditions at each boundary (non-slip at $y=-1$ and symmetry conditions at $y=0$). So matching the interior solutions across the critical layer requires 4 (complex) conditions to determine 4 complex constants. Since the problem is linear and so the amplitude and phase are indeterminate, one of these conditions instead sets the (complex) frequency $c$ for given (real) $\lam$ (in the case of the neutral  curve the determination for $\Im m(c)=c_i$ is replaced by that for $\lam$ to make $c_i=0$). However, there is an extra (complex) unknown in the jump conditions (\ref{v_jump})-(\ref{D3v_jump}), $\alpha_1$, which means a 5th (complex) condition is needed.

In practice, it is convenient to choose $\hv_0=-1$ to set the amplitude and phase of the eigenfunction thereby upping the matching requirement to  6 complex equations (for the 4 complex constants specifying the two outer solutions, $c$ and $\alpha_1$).
An extra equation comes from now being able to impose the  behaviour of $v$ 
\begin{align}
v^{-} +\del \Pi \alpha_1 & = -1 +(-\del \Pi+\tfrac{1}{2}\del^2 \Om -\tfrac{1}{6} \del^3 \Upsilon) [ \,\log |k \Us \del| -\half i \pi \,] +\del \Pi-\tfrac{3}{4} \del^2\Om+\tfrac{11}{36} \del^3 \Upsilon,   \label{v-}\\
v^{+} -\del \Pi \alpha_1  & = -1 +( \del \Pi +\tfrac{1}{2}\del^2 \Om+\tfrac{1}{6} \del^3 \Upsilon) [\, \log |k \Us \del| +\half i \pi \,]-\del \Pi-\tfrac{3}{4}\del^2 \Om-\tfrac{11}{36} \del^3 \Upsilon  \label{v+} 
\end{align}
as the critical layer is approached from either side ($v^{\pm}=v(y^* \pm \del)$). The final extra equation comes from also doing the same for $Dv$
\begin{align}
Dv^{-} -\Pi \alpha_1      & = ( \Pi-\del \Om+\half \del^2 \Upsilon  ) [ \,\log |k \Us \del| -\half i \pi \, ]+\del \Om-\tfrac{3}{4} \del^2 \Upsilon, \label{Dv-} \\
Dv^{+} -\Pi \alpha_1      & =( \Pi+\del \Om +\half \del^2 \Upsilon )[ \,\log |k \Us \del| +\half i \pi \,]-\del \Om -\tfrac{3}{4} \del^2 \Upsilon.  \label{Dv+} 
\end{align}
which crucially does not introduce any new unknown constants. These 4 conditions together with the jump conditions (\ref{D2v_jump}) and (\ref{D3v_jump}) give the required 6 conditions for the 4 unknowns specifying the interior solution below and above the critical layer, the two real numbers,  $c_r$ and $\lam$, and the complex constant $\alpha_1$.


%
%
\begin{table}
\begin{center}
\begin{tabular}{@{}lrrrccrr@{}}
  & $\delta$ \hspace{0.25cm} & \hspace{0.5cm}&    $c_r$ \hspace{0.5cm}  & 
  $k\Lambda_{-}$ &\hspace*{1cm} & $c_r$  \hspace*{0.75cm}&  $k\Lambda_{+}$  \hspace*{0.25cm} \\ \hline
   &                 &                        &  &                                             &                                  &                  &                                               \\
$k=0.1$  \hspace{0.5cm}   &   $32 \times 10^{-4}$     & & 0.99855267713    &       3.446546   & &         &                   \\
            &  $8 \times 10^{-4}$      & & 0.99844946362   &      3.400230    & &          &                \\
            & $2 \times 10^{-4}$      & & 0.99842217083    &      3.389936    & &          &                \\
            & $*1\times  10^{-4}$     & & 0.99841758893    &      3.388332     &&    0.99995720674    &         9.856305  \\
            & $*1\times  10^{-5}$     & & 0.99841347771     &      3.386965     &&    0.99995700930   &        9.823193   \\
            & $*1\times  10^{-6}$     & & 0.99841306978    &      3.386842     &&    0.99995698930   &        9.819909     \\  
            & $ *3\times 10^{-7}$     & & 0.99841304032   &      3.386833      &&    0.99995698769   &        9.819643     \\
            &                         & &                  &                   &&                     & \\
            &  0          & & 0.99841302769 & 3.386829 && 0.99995698700 & 9.819529\\
            &                    & &                   &                   &&                    &    \\
 $k=1.1$    & $32 \times 10^{-4}$   &&   0.99923583992    &  4.525599       & &     0.99988318362     &      8.328840  \\
            & $8 \times 10^{-4}$    &&   0.99922208240    &  4.519946       & &     0.99986369555     &     7.823761   \\
            & $2 \times 10^{-4}$    &&   0.99921892090    &  4.520468       & &     0.99985801319      &     7.706180  \\
            & $*1\times  10^{-4}$   &&   0.99921842343    &  4.520720       & &     0.99985702092     &     7.686784   \\
            & $*1\times  10^{-5}$   &&   0.99921799797    &  4.521053       & &     0.99985611212      &     7.669353   \\
            & $*1\times  10^{-6}$   &&   0.99921795926    &  4.521106       & &     0.99985602002    &     7.667615     \\  
            & $*1\times 10^{-7}$    &&   0.99921795576    &  4.521113       & &     0.99985600515    &      7.667345    \\
            &                         & &                  &                   &&      & \\ 
            &  0          & & 0.99921795537 & 4.521114 && 0.99985600350 & 7.667315\\[6pt]                               
\end{tabular}
\end{center}
\caption{\label{Table4} $c_r$ and $k\Lambda$ found by asymptotic matching on the neutral curve for  $k=0.1$ and $1.1 $ with various choices of $\delta$ ($\pm$ indicates upper and lower parts of the neutral curve and * results computed using quadruple rather than double precision).  The quadruple precision calculations were done using the multi-precision extension package `Advanpix' for Matlab.  The `$0$' predictions are found by applying Richardson extrapolation to the last two entries assuming a leading $O(\del)$ error. All four cases compare well with the numerical predictions marked by `$\infty$' in Table \ref{Table3}.   }
\end{table}

In practice,  we impose the conditions (\ref{D3v_jump}) - (\ref{Dv+}) to specify the velocity fields and $\alpha_1$,  and then  find (real) $c$ and $\lam$ using Newton-Raphson on the remaining  $D^2v$ jump condition (\ref{D2v_jump}).  The matching is not surprisingly quite delicate because up to the third derivative in $v$ has to be matched, there is singular behaviour {\em and} the critical layer can get very close to the centreline which imposes the condition $\del \ll \sqrt{1-c}$.   Invariably $\del$ needs to be smaller than $10^{-4}$  to see convergence which,  since terms up to $O(\del^3)$ need to be resolved,  requires quadruple precision arithmetic.  Table \ref{Table4} show the results of matching at $k=0.1$, which is away from the neutral curve nose (see inset A of figure 2 of \cite{Khalid21b}), and $k=1.1$ which is close to it.  
For the upper neutral curve at $k=0.1$ where $c=0.99957$,  the critical layer  is so close to the centreline that only matching with quadruple precision is possible.

Figure \ref{error} plots the error in estimating the (lower neutral curve) limiting values  of $c$ and $\Lambda$ at $k=1.1$ via the numerical solution taking $W \rightarrow \infty$  and the asymptotic matching approach taking $\delta  \rightarrow 0$.  The asymptotically-matched outer velocity fields over $y \in [-1, y^*-\delta] \cup [y^*+\delta,0]$  compare excellently with the full numerical solution shown in  figure \ref{fig1} - see figure \ref{asymptotic}.  Reducing $\delta$ from $2 \times 10^{-4}$ to $5 \times 10^{-6}$ shows the  singularity in $\Im m(Dv)$ at the critical layer when the phase of the eigenfunction is set by $Dv=1$ at the midplane.  The numerically-computed stress field at $W=32,000$ is compared in figure \ref{stress} with the matched outer solution (using $\delta=2 \times 10^{-4}$) and the leading inner asymptotic solution, again with excellent agreement.

The key realisation  from the asymptotic analysis is that the structure of the critical layer is built upon a non-vanishing cross-stream velocity there. This is reflected in the fact that everything scales with it - specifically $\hv_0$ in the expansion (\ref{v_crit}). For example, the 3 matching constants in (\ref{constants}) are proportional to $\hv_0$: if these are zero, the critical layer has no effect on the outer solutions. The cross-stream velocity, however,  must vanish at the midplane by the symmetry conditions and so there has to be an $O(1)$ outer layer between the critical layer and the centreline to bring about this adjustment (derivatives in the outer regions are $O(1)$ and $\hv_0=O(1)$ to set the normalisation of the eigenfunction). This explains why the critical layer cannot approach the centreline as $W\rightarrow \infty$ or equivalently why $c$ converges to a finite value which is not 1. It remains unclear why  this finite value is numerically so close to 1 (e.g. $1-c=4.3\times 10^{-5}$ for $k=0.1$ in Table \ref{Table3}) but the plausible hypothesis is that the critical layer can only manifest in a low shear region compared to the rest of the domain. This is probed a little in \S 6 below but first we give a discussion on the instability mechanism.

%
%

%
%
\begin{figure}
\centering
\scalebox{0.37}[0.37]{\includegraphics{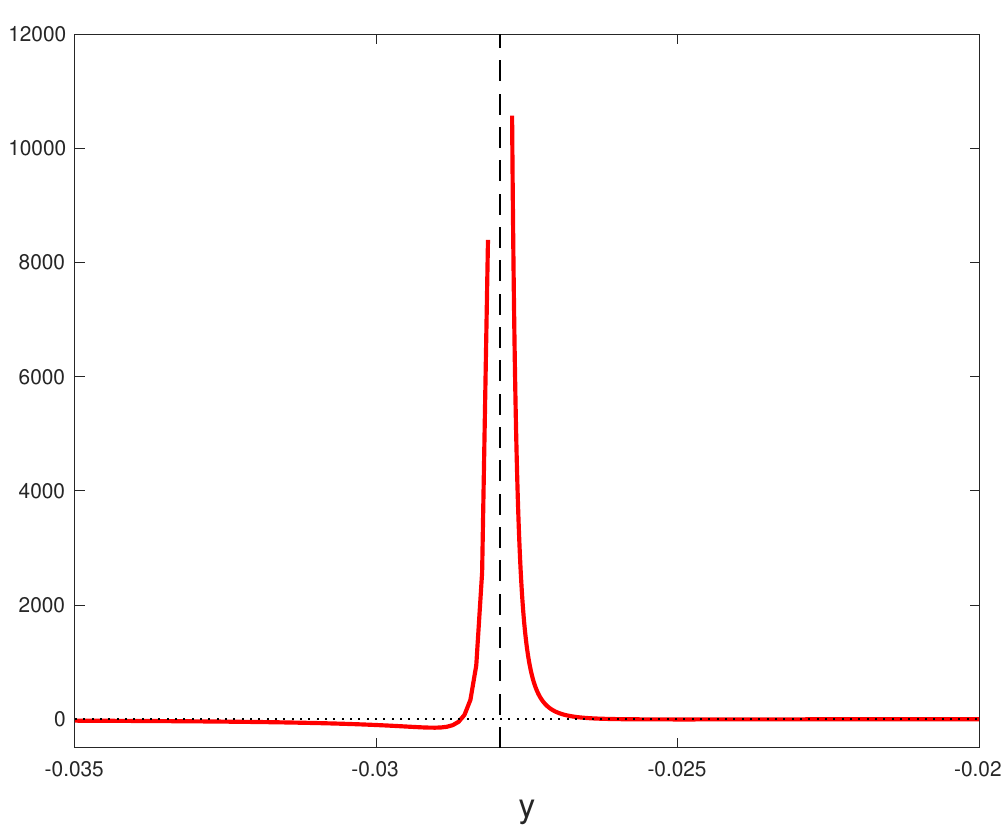}}
\scalebox{0.37}[0.37]{\includegraphics{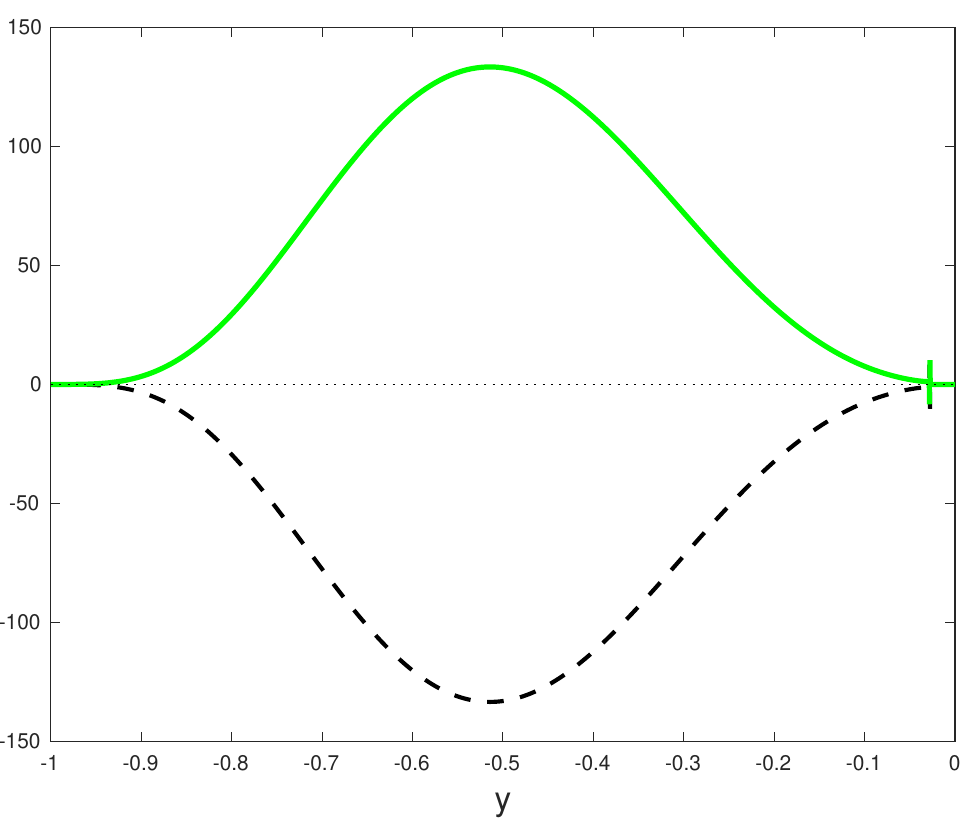}}\\
\caption{\label{balances} Left: $P(y)$ plotted against $y$ for a solution ($k=1.1$, $\Lambda=4.520468$ and $c=0.99921892$) matched across the critical layer using $\delta = 2 \times 10^{-4}$ (the position of the critical layer is shown as the vertical dashed black line). Right: $C(y)$  (green) and  $D(y)$ (black dashed) plotted against $y$ for the same matched solution. These figures show that the velocity field is driven locally by the critical layer whereas the polymer perturbation is  driven globally by the gradual relaxation of the streamline distortion caused by the critical layer.}
\end{figure}

\section{Instability Mechanism}

The asymptotic analysis above separates the `inner' solution in the critical layer  from the `outer' solution, allowing scrutinization  of how the instability manifests in the latter. The role of the critical layer is then viewed as generating `energising' internal conditions for the outer solution (or boundary conditions for the two parts of the outer solution). Before proceeding in this manner, we simplify  the outer equations
%
%
by rewriting some terms using the streamline displacement $\phi:=v/ik(U-c)$ instead of $v$ following \cite{Rallison95} to get
\begin{align}
0 & = -ikp+(D^2-k^2) u -ik T_{11} D \phi ,           \label{mod_x}\\
0 & =-Dp+(D^2-k^2) v  -k^2 T_{11}   \phi,            \label{mod_y}\\
0 & = iku+Dv,                                        \label{mod_incompress}\\
\tau_{11}  &= -2T_{11}D \phi -\phi D T_{11},          \label{mod_t11}\\
\tau_{12}  &= ik T_{11} \phi.                         \label{mod_t12}
\end{align}
While the above analysis shows that $v$ is continuous across the critical layer, $\phi \sim 1/(U-c)$ as the critical layer is approached. Hence the streamline displacement is maximal there and the question is how this drives the polymer field which must in turn offset the viscous dissipation in the inertialess momentum equation.

\begin{figure}
\centering
\scalebox{0.37}[0.37]{\includegraphics{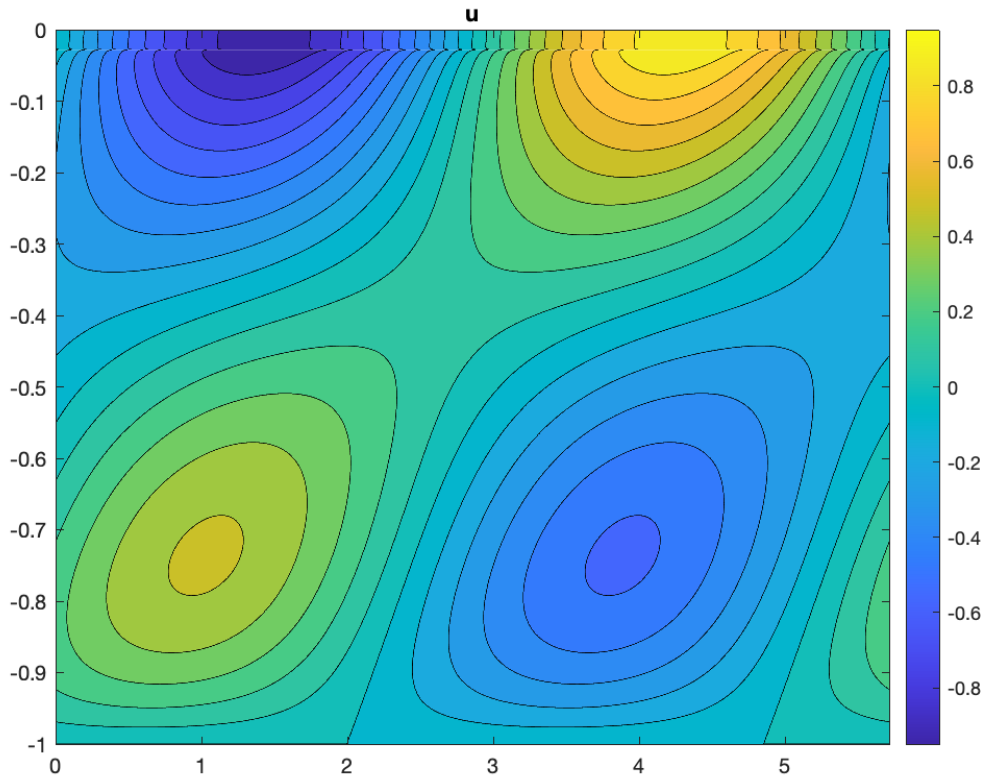}}
\scalebox{0.37}[0.37]{\includegraphics{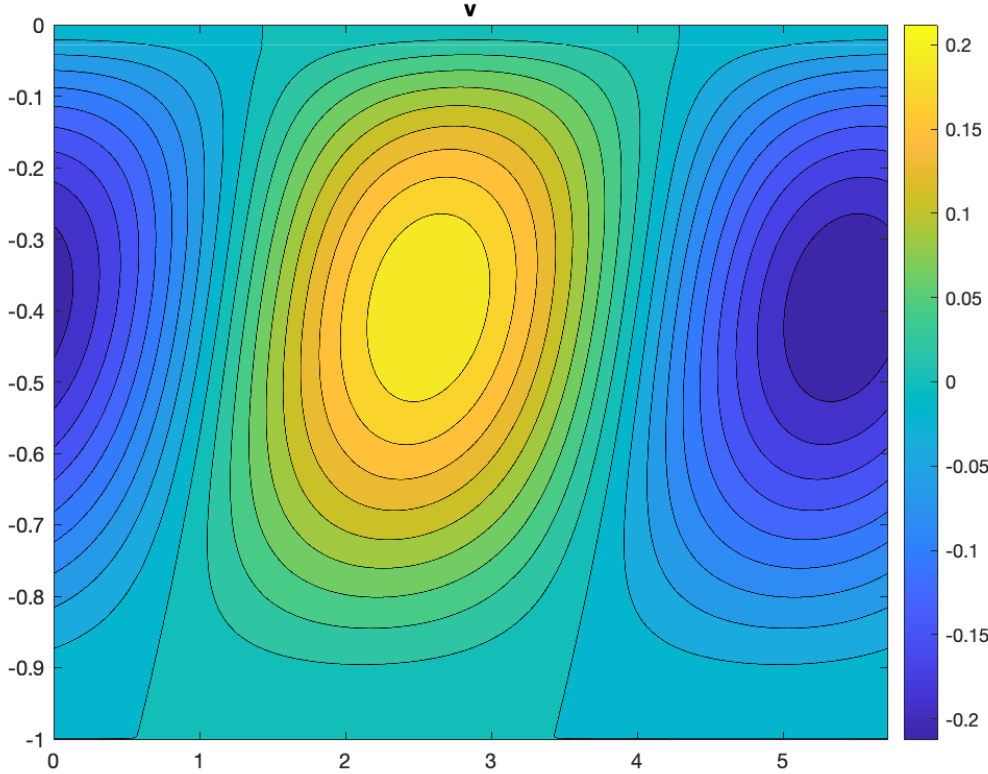}}\\
\scalebox{0.37}[0.37]{\includegraphics{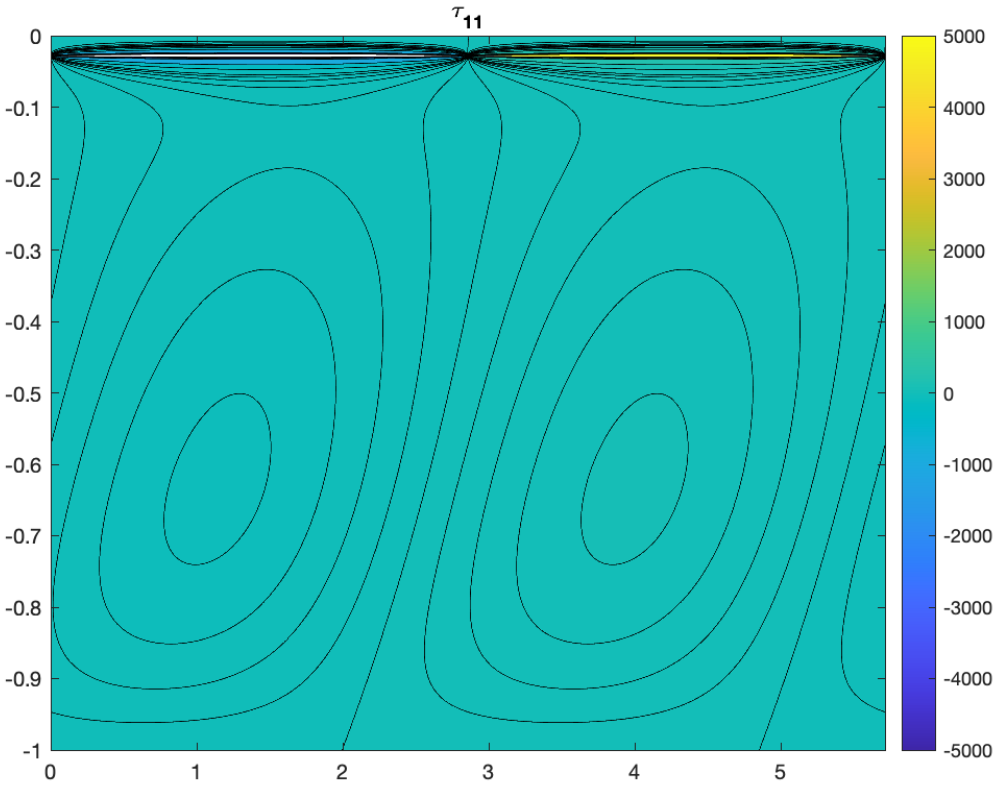}}
\scalebox{0.37}[0.37]{\includegraphics{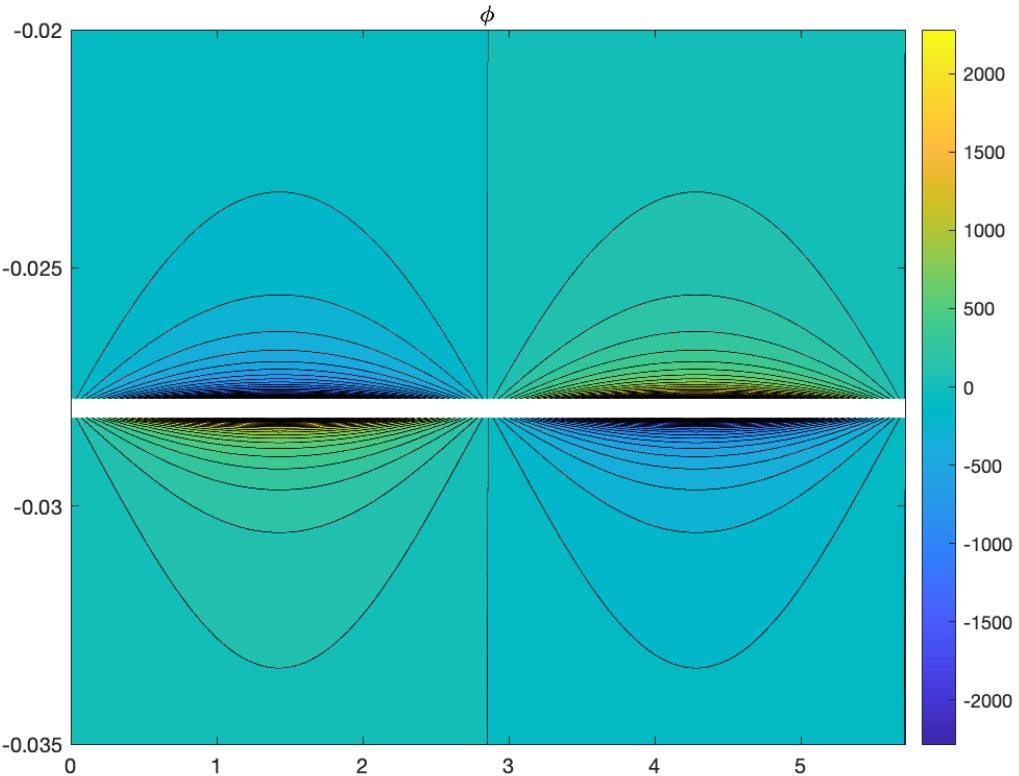}}\\
\caption{\label{typical} Contour plots of the outer solution corresponding to $k=1.1$, $\Lambda=4.520468$ and $c=0.99921892$ matched across the critical layer using $\delta = 2 \times 10^{-4}$. Top left $u$, top right $v$, bottom left $\tau_{11}$ and bottom right $\phi$. The last plot focusses on the area around the critical layer (the white gap is just the excluded critical layer region  $[y^*-\delta, y^*+\delta]$). For $u$ and $v$, 20 equally spaced contour levels between the maximum and minimum values are used. For $\tau_{11}$, the contour levels are  $\pm[0,10,20,30,40,50,60,100,200,1000,2000,5000]$ and 50 equally spaced levels are used for $\phi$ with all the contours essentially contained in the shown area around the critical layer.. }
\end{figure}

Approaching the neutral curve, $c_i \rightarrow 0$, means that $\phi$ is exactly out of phase with $v$ so that $-k^2 T_{11}\phi$ can do no work in energising $v$ in (\ref{mod_y}), that is,
\beq
\frac{k}{2 \pi} \int^{2\pi/k}_0 \Re e[ v e^{ik(x-ct)} ] \Re e [-k^2 T_{11} \phi e^{ik(x-ct)} ]\, dx =0.
\eeq
This means the viscous dissipation in the cross-stream variable $v$ can only be offset by the pressure term and the polymer stress driving of the velocity field must occur in (\ref{mod_x}). To confirm this, the power input
\beq
P(y) := \frac{k}{2 \pi} \int^{2\pi/k}_0 
\Re e [ u e^{ik(x-ct)} ] \Re e [ -ik T_{11} D \phi e^{ik(x-ct)} ]
\, dx
\label{P}
\eeq
is plotted in figure \ref{balances} which clearly shows that the polymer stress energising of $u$ is localised at the critical layer. The specific solution used here is shown in figure \ref{typical}.

%
%

The polymer stress equations are particularly simple when written in terms of the streamline displacement and interpretable. The first term on the right of (\ref{mod_t11}) represents the increase in the streamwise-normal stress due to streamline compression (in $y$), the second term represents the maintenance of the initial basic stress on displaced streamlines, and the third term (the  the RHS of (\ref{mod_t12})\,) is simply the generation of tangential polymer stress due to tilting of the base polymer stress lines \citep{Rallison95}. Multiplying (\ref{mod_t11}) by $ik(U-c)$ recovers the time-derivative and advection terms on the LHS of (\ref{mod_t11}) and thereby recovers the proper driving terms on the RHS,
\beq
ik(U-c) \tau_{11}  =   -2ik(U-c)T_{11}D\phi-ik(U-c)\phi DT_{11}. 
\label{preD}
\eeq
Their energising effects,
\begin{align}
C(y) &:= \frac{k}{2 \pi} \int^{2\pi/k}_0 
\Re e [ \tau_{11} e^{ik(x-ct)} ] \Re e [ -2ik(U-c)T_{11}D \phi e^{ik(x-ct)} ]
\, dx  \label{D1}\\
\& \quad 
D(y) &:= \frac{k}{2 \pi} \int^{2\pi/k}_0 
\Re e [  \tau_{11} e^{ik(x-ct)} ] \Re e [ -ik(U-c)\phi DT_{11} e^{ik(x-ct)} ]
\, dx
\label{D2}
\end{align}
are shown in  figure (\ref{balances}). Both terms are global and barely register the critical layer with streamline  $C$ompression causing the streamwise-normal stress to increase ($C(y)$) while $D$isplacement across the base shear field ($D(y)$) works negatively to balance it on the neutral curve. It is worth remarking that the reintroduction of the factor $(U-c)$ into (\ref{preD}) is responsible for desensitizing $C(y)$ to the critical layer as otherwise a significant component - $T_{11}D\phi$ - is the same as that in $P$.

The mechanism of the instability can therefore be seen as one in which the  critical layer acts like a pair of `bellows' periodically sucking the flow streamlines together - see $\phi(\pi/2,y)$ in figure \ref{typical} - and then blowing them apart - see $\phi(3\pi/2,y)$  same figure. This amplifies the base streamwise-normal stress field which in  turn exerts a streamwise stress on the flow locally at the critical layer. The streamwise flow drives the cross-stream flow through continuity which then intensifies the critical layer closing the loop.

The one outstanding question is why the critical layer has to be so close to the centreline as $W \rightarrow \infty$. The asymptotic analysis above indicates that the shear at the critical layer needs to be $O(W^0)$ as $W \rightarrow \infty$ but can tell us nothing about the size of this  $O(W^0)$  number relative to 1. In particular, given the complicated analysis, it is still not clear why the instability does not manifest in plane Couette flow. To help answer this, we conduct some simple experiments in the next section.

%
%

\section{Moving towards plane Couette flow}

The complexity of the matching analysis means it is difficult to discern the importance of $U_*^{''}$ or the size of $\Us$ (e.g. just look at the 3 constants that emerge in (\ref{constants})\,) despite our best efforts to keep them separated. So here, we perform a series of numerical experiments exploring the effect of small changes which would bring the channel flow closer to plane Couette flow (pCf). To keep things manageable, these experiments concentrate on studying how the lowest $W$ point on the neutral curve, $W_{min}$, varies as the problem is changed slightly. Four experiments are undertaken in which this minimum point is tracked as a homotopy parameter $\lambda$ is reduced from 1 which is the channel flow problem studied above where $W_{min}=974$. These are as follows.\\[3pt]
\begin{itemize}
\item Expt. 1 explores the importance of $U^{''}$ by (artificially) changing it while keeping $U^{'}$ fixed. Specifically $U^{''}:=-2\lambda$  everywhere so $U^{''}$ can be reduced without changing $U=1-y^2$ or $U^{'}=-2y$ in the code.
\item Expt 2 explores the effect of changing the boundary condition at the midplane towards a solid boundary to mimick the monotonic increase in $U$ across the domain of pCf. The boundary condition at $y=0$ is set to $\lambda D^2v+(1-\lambda)Dv=0$  so $\lambda=1$ corresponds to the stress-free/symmetry conditions considered above and $\lambda=0$ to a non-slip solid wall.
\item Expt. 3 explores the effect of increasing the minimum shear across the domain $y \in [-1,0]$. The base flow is set to $U:=\lambda (1-y^2)+(1-\lambda)y$ which mixes in pCf in a way  to gradually generate a (minimum) non-zero shear $=1-\lambda$ at the midplane.
\item Expt 4 explores the effect of moving the $U^{''}=0$ point into the interior. The base flow is set to $U:=\lambda (1-y^2)-(1-\lambda)y$ which mixes in pCf in a way to move the zero-shear point at $y=-(1-\lambda)/2\lambda$  away from the midplane and into the interior $y \in [-1,0]$.\\[3pt]
\end{itemize}

\begin{figure}
\centering
\scalebox{0.35}[0.35]{\includegraphics{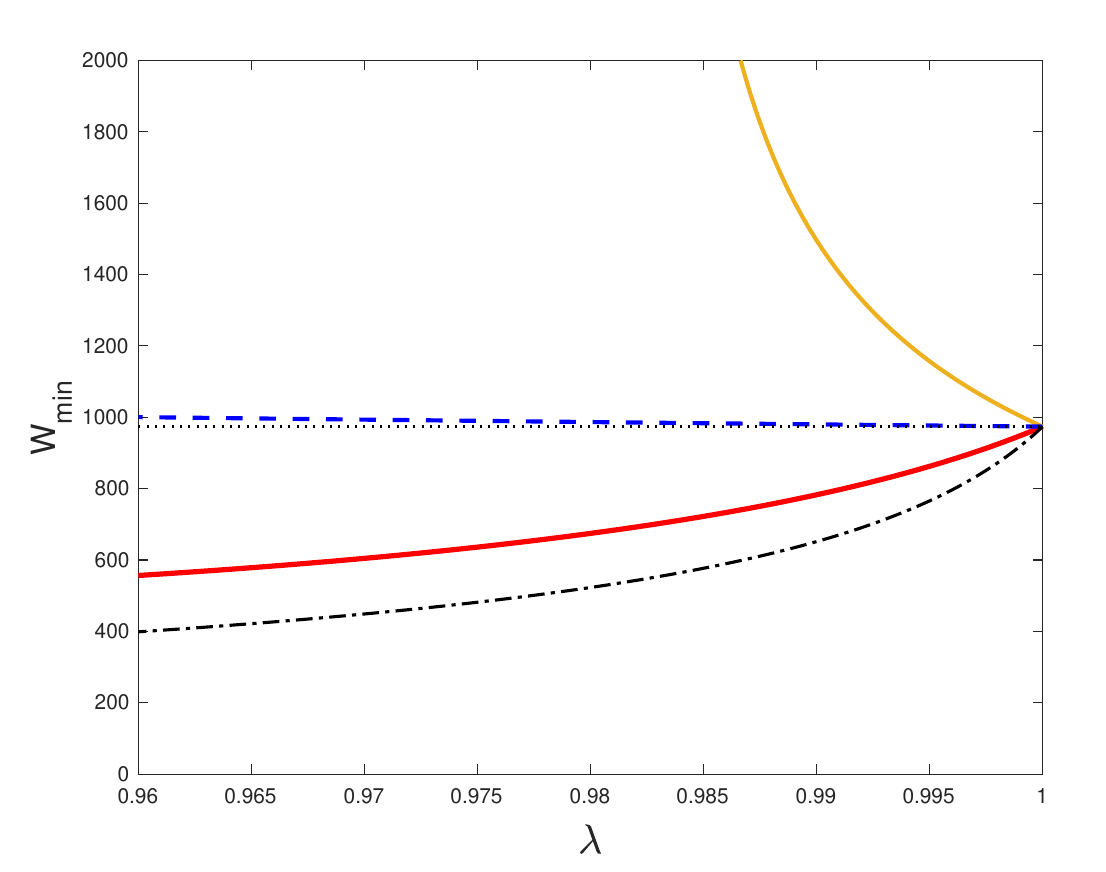}}
\scalebox{0.35}[0.35]{\includegraphics{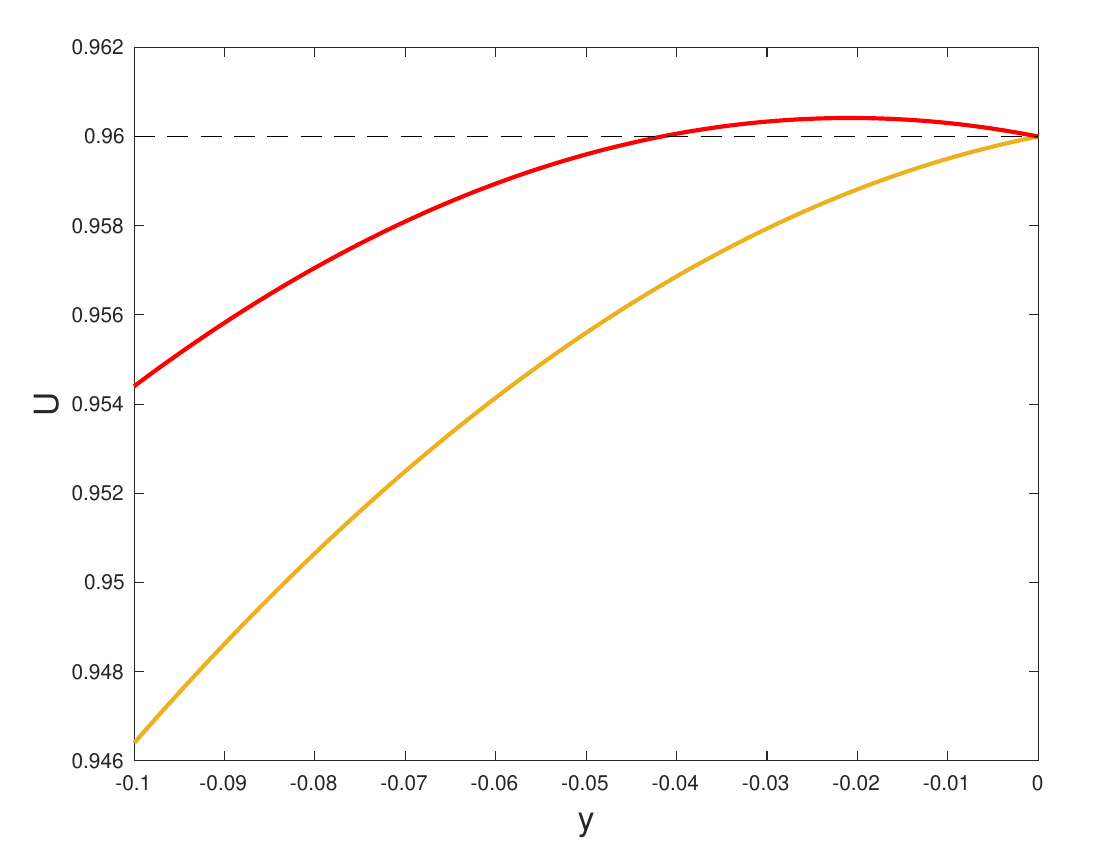}}
\caption{Left: the effect on $W_{min}$ of decreasing $\lambda$ from 1 to 0.96 (the common point on $\lambda=1$ is the undisturbed value of $W_{min}=974$): Expt. 1 black dash-dot line; Expt. 2 blue dashed line; Expt. 3 yellow solid line; Expt. 4 red solid line. Right: the base flow profile $U:=\lambda(1-y^2)+\lambda y$ in Expt. 3 (yellow line) and $U:=\lambda (1-y^2)-\lambda y$ in Expt. 4 (red line) for $\lambda=0.96$ showing  how the former eliminates the zero-shear point whereas the latter moves it into the interior (notice only a small part of the flow domain is being shown close to the midplane. In all cases the problem is solved over $y \in [-1,0]$) 
\label{expts} 
}
\end{figure}

The results are shown in figure \ref{expts}. The  effect of changing the boundary conditions (blue dashed line) at the midplane is minimal and reducing $U^{''}$ (black dash-dot line) is {\em de}stabilizing.  The presence of vanishing shear, however, seems crucial: removing it (yellow line) quickly stabilizes the instability whereas moving it from the midplane (red line) is destabilizing. The plausible conclusion is that the instability needs small shear relative to the rest of the domain to nucleate. This is certainly absent in pCf.
 
%
%

%
%

%
%
\begin{figure}
\centering
\scalebox{0.35}[0.35]{\includegraphics{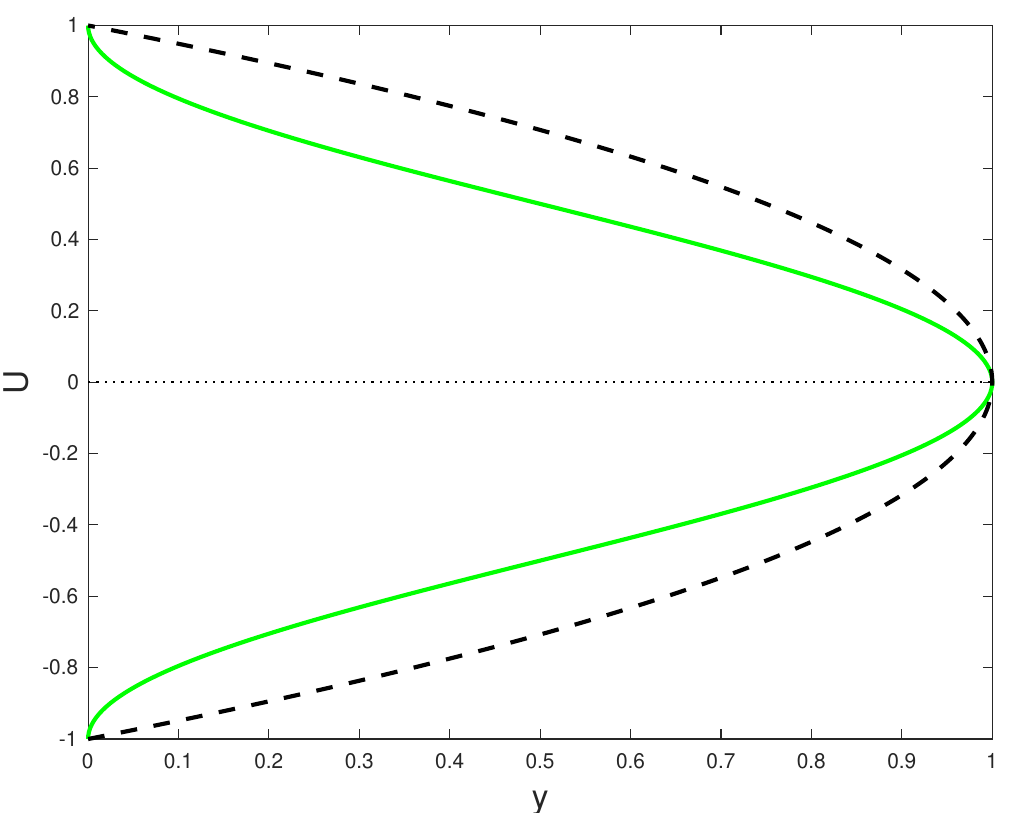}}
\scalebox{0.35}[0.35]{\includegraphics{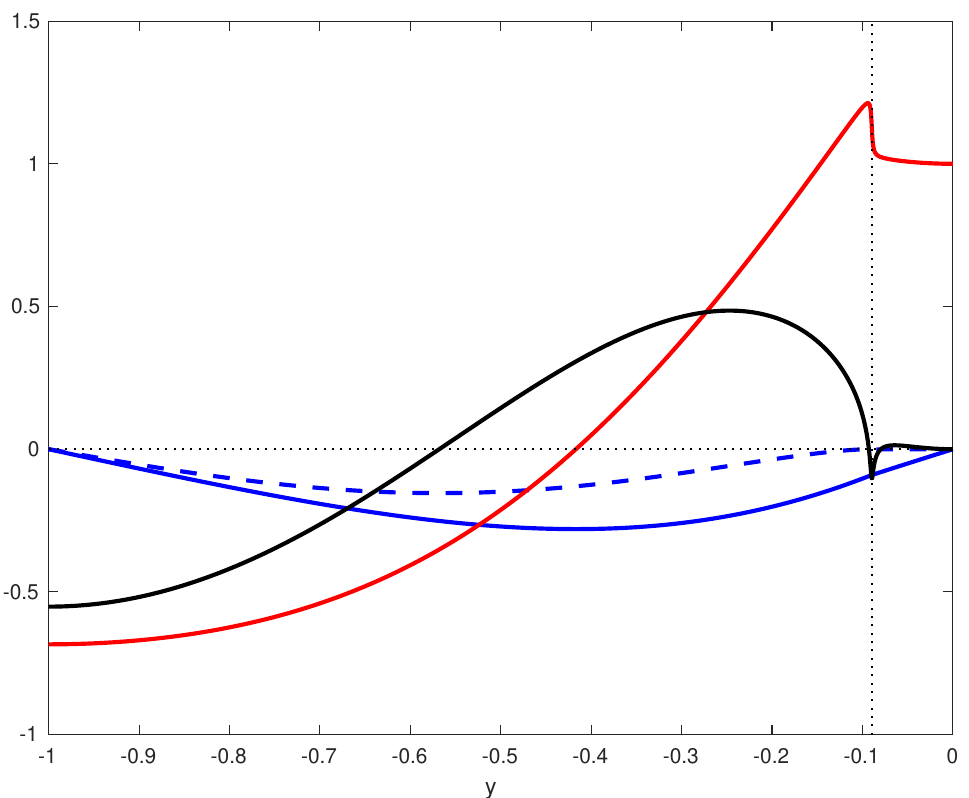}}\\
\scalebox{0.35}[0.35]{\includegraphics{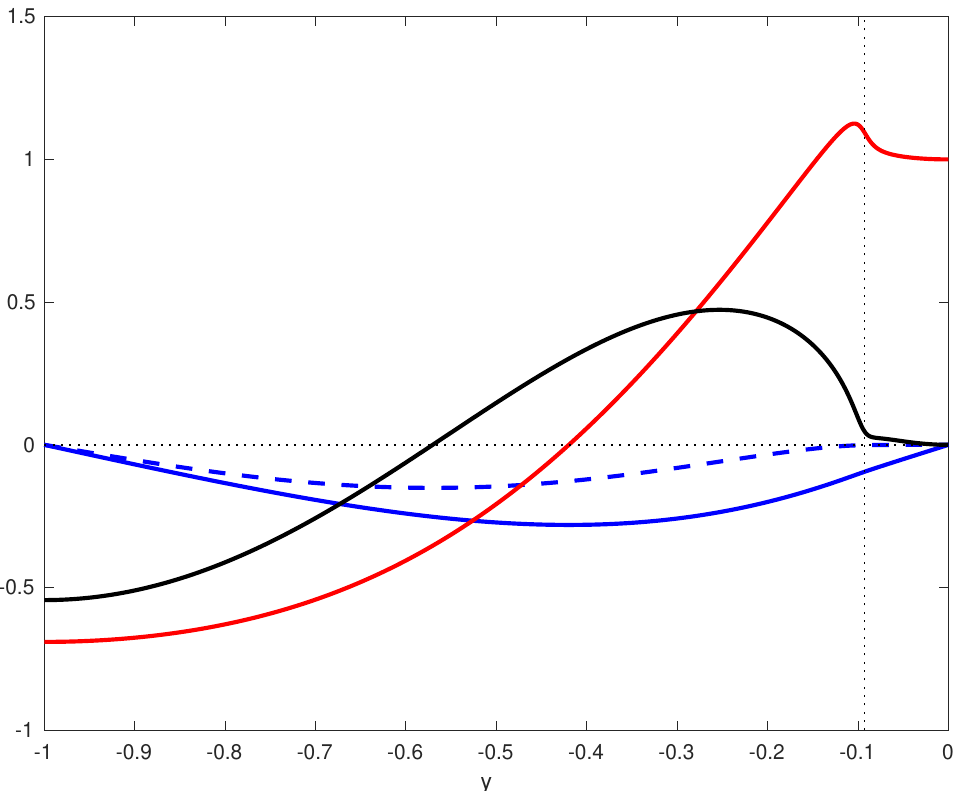}}
\scalebox{0.35}[0.35]{\includegraphics{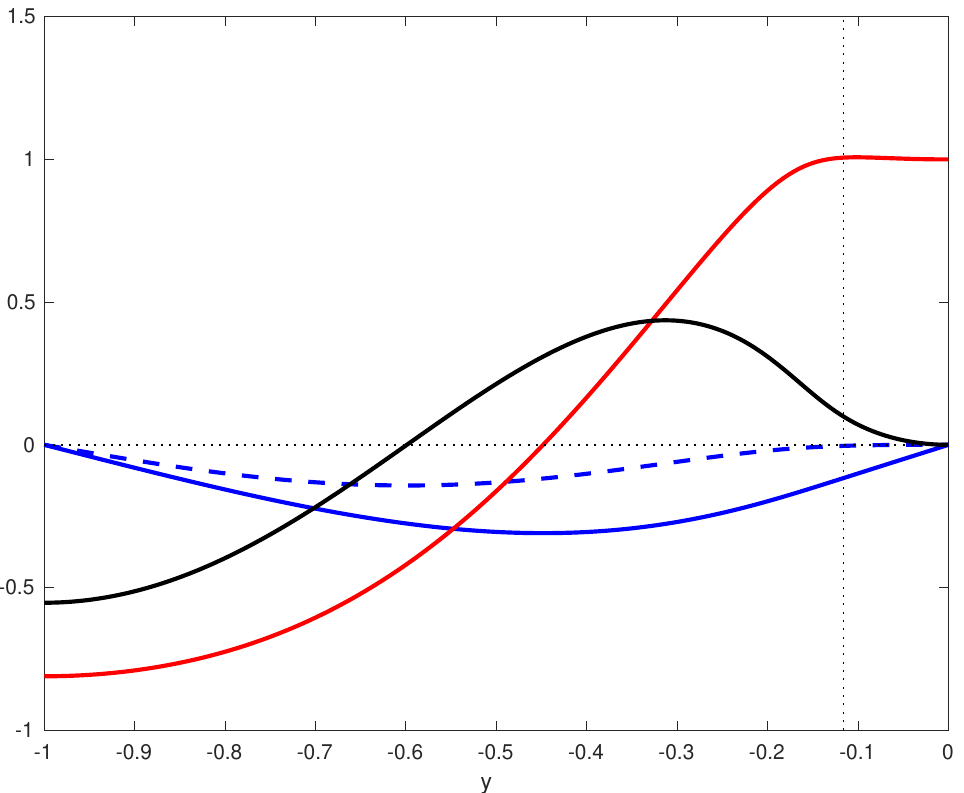}}\\
\caption{\label{vKf} Top left: Viscoelastic Kolmogorov base flow as given in (\ref{vKf_base}) in green and channel flow $(1-y^2)$ in dashed black. Neutral vKf eigenfunctions for $k=1$ and 
$(W,\beta,c)=(2000, 0.99954, 0.98029)$ (top right);  
$(W,\beta,c)=( 200, 0.99530, 0.97897)$(bottom left) 
and 
$(W,\beta,c)=(  20, 0.93972, 0.96726)$(bottom right). $v$ is blue (real/imaginary parts solid/dashed respectively),  real part of $Dv$ is red and imaginary part is black.}
\end{figure}

\section{Viscoelastic Kolmogorov flow}

Finally, the similarity of the base flow shape in viscoelastic Kolmogorov flow (vKf) to that in channel flow suggests that the viscoelastic linear instability found there \citep{Boffetta05} could be the centre mode instability of \cite{Garg18,Khalid21a,Khalid21b}. It is straightforward to confirm this by renormalising the base flow in vKf to the form
\beq
U = \tfrac{1}{2} (1+\cos \pi y ) 
\label{vKf_base}
\eeq
(to most closely match $(1-y^2)$ - see Figure \ref{vKf}) and considering disturbances which are periodic over $y\in[-1,1]$ {\em and} have the same symmetry (\ref{symmetry}) as the centre mode instability about $y=0$. These properties actually imply that the disturbance satisfies stress-free boundary conditions at $y=\pm 1$ and so all the numerical codes developed for channel flow can trivially be reapplied to vKf by just i) changing $U(y)$ and ii) imposing stress-free boundary conditions on the perturbation at $y=-1$.

Using a shooting code which identifies the neutral curve at a given $k$ and $W$ after a guess for $\beta$ and $c$ (based on the channel flow solution) quickly identifies a neutral vKf eigenfunction for $(W,k)=(2000,1)$ at $(\beta,c)=(0.99953538, 0.98029137)$. Reducing $W$ to $200$, keeping $k=1$ and using the values of $\beta$ and $c$ found at $W=2000$ as initial guesses, the neutral eigenfunction at $W=200$ converges easily to $(\beta,c)=(0.99530288, 0.97896974)$. Repeating this procedure, reducing $W$ from $200$ to $20$, again converges smoothly to  $(\beta,c)=( 0.93972375, 0.96725546)$: see figure \ref{vKf}. The  similarity of the neutral eigenfunctions in figure \ref{vKf}  to those in channel flow (modulo the different boundary conditions at $y=-1$) is striking and suggests that \cite{Boffetta05} were actually the first to find the centre mode instability in rectilinear viscoelastic flow. To further support this conclusion, \cite{Berti08, Berti10} see `arrowhead' solutions when tracking their vKf instability to finite amplitude (e.g. see figures 7 and 8 in \cite{Berti10}) just like those found in channel flow originating from the centre mode instability \cite{Page20,Buza22b, Beneitez23b}.

Lastly, it is also worth remarking that $W$ can be reduced down to at least $2.1$ at $k=1$ where $(\beta,c) =  0.035543817, 0.82144631)$ in vKf. These $W$ and $\beta$ are an order of magnitude smaller than those required for centre mode instability in channel flow presumably because of the rigid boundaries present. 

%
%

\section{Discussion}

%
%

We first summarise the findings of the paper.  The first part of these concern the  $Re \rightarrow \infty$ asymptotics of the upper (\S3.1) and lower branches (\S3.2) of the centre-mode neutral curve in the $Re$-$W$ plane for viscoelastic channel flow. 

Along the upper branch
\beq
W \sim Re^{1/3}, \quad k \sim Re^{1/3}, \quad 1-c \sim Re^{-2/3}
\eeq
with numerical coefficients given in (\ref{upperscalings}) for $\beta=0.9$. These scalings are equivalent to $Re \sim O(E^{-3/2}$), $k \sim O(E^{-1/2})$ and $1-c \sim O(E)$ as the elasticity number $E:=W/Re \rightarrow 0$ consistent with figure 11 in \cite{Khalid21a}.

Along the lower branch,
\beq
W \sim Re, \quad k \sim \frac{1}{Re}, \quad c = O(1)
\eeq
with numerical coefficients computed for $\beta=0.9, 0.98$ and $0.994$ given in Table \ref{Table2}. These lower branch scalings are apparent in figure 13 of \cite{Khalid21a} (see also their figure 18).  

 The second part of the findings described in \S4 concern the inertialess limit of viscoelastic channel flow. By $\beta=0.994$ as $\beta$ increases, the lower branch has swung sufficiently clockwise in the $Re$-$W$ plane to cross the $Re=0$ axis (see figure \ref{ub_lb_curves}). This reveals the existence of an inertialess ($Re=0$) centre mode instability and the $W \rightarrow \infty$ asymptotic problem in the ultra-dilute limit where $W(1-\beta)=O(1)$ is then treated.  A matched asymptotic analysis is performed in which a critical layer region is resolved sufficiently to extract matching conditions, (\ref{v_jump})-(\ref{D3v_jump}), to connect up outer regions either side. Interestingly, the outer problem is 4th order as opposed to the usual 2nd order problem for the Orr-Sommerfeld problem and so requires matching conditions all the way down to the 3rd order derivative in the cross-stream velocity. This leads to a particularly delicate matching procedure (\S4.8) where the matching conditions need to be resolved to third order in the small matching parameter and  quadruple precision is needed to make contact with numerical solutions. The completed analysis is successful in revealing, again unlike the Orr-Sommerfeld problem, that
\beq
1-c = O(1) \qquad {\rm as}\quad W \rightarrow \infty 
\eeq
This $O(1)$ number can be deceivingly small compared to 1 (e.g. $4.3\times 10^{-5}$ for $k=0.1$ in Table \ref{Table3}) but nevertheless remains finite as $W \rightarrow \infty$. That this has to be so is clear from the structure of the critical layer that  is built around an $0(1)$ cross-stream velocity which has to be brought to zero at the midplane by an $O(1)$ outer region separating  the critical layer from it. Quite why $c$ has to be so close to 1 or equivalently why the critical layer sets up in a region of small shear is unclear (and unknowable from the asymptotic analysis). Some simple numerical experiments (\S6) suggest that the lack of this small shear region is the likely reason the instability does not manifest in plane Couette flow. 

The asymptotic analysis also clarifies that the instability mechanism (\S5) is one in which the critical layer acts like a pair of `bellows' periodically sucking the flow streamlines together and then blowing them apart  (see figure \ref{typical}). This amplifies the streamwise-normal polymer stress field which in  turn exerts a streamwise stress on the flow locally at the critical layer. The streamwise flow drives the cross-stream flow by continuity which then intensifies the critical layer closing the loop.

Finally in \S7, a connection is made between the centre mode instability of channel flow and an earlier linear instability found in viscoelastic Kolmogorov flow by \cite{Boffetta05}. The fact that the instability in Kolmogorov flow was discovered at much lower $Re$ and $W$ to the extent it was viewed as a purely `elastic' instability disguised its connection to the work of \cite{Garg18}. They worked at  $Re=O(100)$-$O(1000)$ and $W \gtrsim 20$ in a very different geometry and so viewed their instability as `elasto-inertial' in origin. It is clear in hindsight that the apparent difference in the regimes is more a function of the boundary conditions - a solid wall in the pipe verses periodicity in Kolmogorov flow - than any deeper dynamical difference as evident in the Newtonian versions of the respective problems ($Re_{crit}=O(10)$ for Kolmogorov flow while $Re_{crit}=5772$ for channel flow).

%
%

The importance of the centre mode instability for elasto-inertial turbulence (EIT) or indeed elastic turbulence (ET) is still an area of much current speculation \cite[e.g.][]{Datta22, Dubief23}. While computations have confirmed that the instability leads to travelling wave solutions dubbed `arrowheads \citep{Page20,Dubief22,Buza22b}, it remains unclear what these lead to via their own bifurcations. Recently \cite{Beneitez23b} have found that the arrowhead solutions coexist with EIT rather breaking down to it.  The situation, however, is slightly clearer at  $Re=0$ (perhaps because the parameter space is one dimension less) where the (2-dimensional) arrowhead solution can become unstable to 3-dimensional disturbances \citep{Lellep23a}. Very recent calculations using a large domain indicate that this instability can lead to a 3-dimensional chaotic state \citep{Lellep23b}.

Further complicating the picture is the very recent emergence of another viscoelastic instability - dubbed `PDI' for polymer diffusive instability - when polymer stress diffusion is present \citep{Beneitez23a, Couchman23, Lewy23}. This is a `wall' mode  which also exists for all $Re$ including $0$ and any shear flow with a solid wall is susceptible.  \cite{Beneitez23a} have already found that PDI can lead to a chaotic 3D state in inertialess plane Couette flow using the FENE-P model. 

Going forward, the challenge is to try to unpick which process of the current contenders - viscoelastic Tollmien Schlichting instability, the centre mode instability or the PDI - triggers EIT and ET in what part of parameter space. This will assist in simulating EIT and ET and ultimately in manipulating those states as required for industrial applications. 

\vspace{1cm{}}
\noindent
Acknowledgements: The authors acknowledge EPSRC support for this work under grant EP/V027247/1.\\[5pt]

\noindent
Declaration of interests: The authors report no conflict of interest.\\[5pt]

%
%

%
%
\appendix

\section{Numerical methods for eigenvalue problem}

Two complementary approaches were developed: a matrix formulation  and a shooting technique.

\subsection{Matrix}
%
%

A generalised eigenvalue code was written to solve for all 6 variables $(u,v,p,\tau_{11},\tau_{12}, \tau_{22})$ building in the symmetry of the unstable eigenfunction around the midline to improve efficiency.  This was done by mapping the lower half of the channel $y \in [-1,0]$ to the full Chebyshev domain $[-1,1]$ so collocation points are concentrated at the wall and the centreline,  and imposing symmetry boundary conditions at the centreline $y=0$ which are $\partial u / \partial y=v=0$ (no b.c.s are imposed on $p$ or the polymer stress $\btau$ anywhere).  The fields are expanded using individual functions which incorporate these conditions: specifically
\beq
\left[ 
\begin{array}{c}
u \\
v\\
p\\
\tau_{11}\\
\tau_{12}\\
\tau_{22}
\end{array}
\right]
= 
e^{ik(x-ct)} \sum_{n=1}^N \left[
\begin{array}{c}
u_n \{\, T_{n+1}(2y+1)+\alpha_n T_n(2y+1)+(\alpha_n-1)T_{n-1}(2y+1)\, \}\\
v_n \{ \, T_{n+1}(2y+1)-T_{n-1}(2y+1) \,\}\\
p_n T_{n-1}(2y+1)\\
t_n T_{n-1}(2y+1)\\
r_n T_{n-1}(2y+1)\\
s_n T_{n-1}(2y+1)
\end{array}
\right]
\nonumber
\eeq
where $T_n(z):=\cos( n\cos^{-1}z)$ is the $n$th Chebyshev polynomial,  $(u_n,v_n,p_n, t_n, r_n, s_m) \in {\mathbb C}^6$ and $\alpha_n:=-4n/(2n^2-2n+1)$ so that $u(x,-1,t)=0=\partial u/ \partial y(x,0,t)$.  A complementary inverse iteration code was also written  which could take an eigenvalue from the generalised eigenvalue code and converge it at much higher resolution  (e.g. Table 1).  This was important as the generalised eigenvalue problem is not well-conditioned with increasing resolution: see the drift in the eigenvalue for $N\gtrsim300$ in Table 1 using {\it eig} in Matlab).  This lack of conditioning gets worse near the neutral curve where an interior critical layer is present.  Inverse iteration treats  exactly the same matrices but is far better conditioned - there is no drift in Table 1 even increasing $N$ to 2000.

\subsection{Shooting}

Two shooting codes were also written based on different integrators.  The first used  RK4 over a uniformly spaced grid (e.g.  50,000 points across [-1,0] in Table 1)  with inbuilt re-orthogonalisation of shooting solutions across the domain and a second used Matlab's ODE15s  with relative and absolute tolerances set at $3\times 10^{-14}$ which did not.  For the solutions sought, re-orthogonalisation was not needed and so the latter, which was more efficient as it has locally adaptive stepping, was used for all subsequent calculations.  The eigenvalue problem is 4th order so the usual shooting code approach takes a guess for the complex phase speed $c$ and searches for the 2 unknown velocity boundary conditions at one wall which mean that the required boundary conditions at the other are  obeyed.  This can be readily adapted to search for the neutral curve directly by setting $c_i=0$ and instead adjusting one (real) parameter of the problem.  Here we chose to vary $\Lambda=(1-\beta)W$ keeping $W$ fixed.   This can be used to recreate the upper inset in figure 2 of \cite{Khalid21b}: see Tables 2  and 3 for sample computations at $k=0.1$ and $k=1.1$.


%
%
\begin{table}
\begin{center}
\begin{tabular}{@{}lrlcrc@{}}
                            & $N$ & \hspace{0.5cm}  &   $c_r$    &\hspace{0.5cm} &   $c_i$ \\
                            & &                           &      &      &          \\
Eigenvalue code &   100    &                             & 0.999608051011  &  &8.2367630455$\times 10^{-5}$\\
                            &   200   &                            & 0.999608009040 &  &8.2751793055$\times 10^{-5}$\\
                            &   300   &                             & 0.999607999400 &  &8.2756092074$\times 10^{-5}$\\
                            &    500  &                             & 0.999608029589 &  &8.2771153285$\times 10^{-5}$\\
                            &  1000   &                            &  0.999608103554  &  &8.2644316360$\times 10^{-5}$  \\ 
                            &  2000 &                             &  0.999607409613   &  &8.3760120903$\times 10^{-5}$\\
                           & &                           &      &      &          \\
Inverse iteration & 100    &                             &0.999608051169   &  & 8.2367612043$\times 10^{-5}$\\
                            &  200  &                             &0.999608007042  &  &8.2751523639$\times 10^{-5}$\\
                            &  300   &                            & 0.999608007115   &  &8.2751700997$\times 10^{-5}$\\
                            & 500   &                            & 0.999608007115   &  &8.2751701045$\times 10^{-5}$\\
                            & 1000   &                            & 0.999608007115   &  &8.2751701039$\times 10^{-5}$\\
                            & 2000  &                           & 0.999608007115    &  &8.2751701041$\times 10^{-5}$\\
                           &             &                                 &   &                                                    \\
Shooting            &ODE15s &                       & 0.999608007115  &   &8.2751701043$\times 10^{-5}$\\
                           &50,000 &                        &0.999608007115  &   &8.2751701040$\times 10^{-5}$\\
                           &              &                         &                               &   &                                       \\[6pt]
\end{tabular}
\end{center}
\caption{\label{Table8}Check of codes with the eigenvalue shown in figure 1(inset) of \cite{Khalid21b}.  The unstable centre mode at $\beta=0.997$, $k=0.75$ and $W=2500$ is shown there with  $c_r \approx 0.9996$ and $c_i \approx 8 \times 10^{-5}$.    Two shooting codes were written based on different integrators.  The first used  RK4 over a uniformly spaced grid (here 50,000 points across [-1,0]) and the second used Matlab's ODE15s  with relative and absolute tolerances set at $3\times 10^{-14}$.  }
\end{table}

\bibliographystyle{jfm}
\bibliography{references}

\begin{thebibliography}{31}
\expandafter\ifx\csname natexlab\endcsname\relax\def\natexlab#1{#1}\fi

\bibitem[Beneitez {\em et~al.\/}(2023{\natexlab{{\em a\/}}})Beneitez, Page,
  Dubief \& Kerswell]{Beneitez23b}
{\sc Beneitez, M., Page, J., Dubief, Y. \& Kerswell, R.~R.} 2023{\natexlab{{\em
  a\/}}} Multistability of elasto-inertial two-dimensional channel flow. {\em
  https://arxiv.org/abs/2308.11554\/} .

\bibitem[Beneitez {\em et~al.\/}(2023{\natexlab{{\em b\/}}})Beneitez, Page \&
  Kerswell]{Beneitez23a}
{\sc Beneitez, M., Page, J. \& Kerswell, R.~R.} 2023{\natexlab{{\em b\/}}}
  Polymer diffusive instability leading to elastic turbulence in plane
  {C}ouette flow. {\em Phys. Rev. Fluids\/} {\bf 8}, L101901.

\bibitem[Berti {\em et~al.\/}(2008)Berti, Bistagnino, Boffetta, Celani \&
  Musacchio]{Berti08}
{\sc Berti, S., Bistagnino, A., Boffetta, G., Celani, A. \& Musacchio, S.} 2008
  Two-dimensional elastic turbulence. {\em Phys. Rev. E\/} {\bf 77}, 055306.

\bibitem[Berti \& Boffetta(2010)]{Berti10}
{\sc Berti, S. \& Boffetta, G.} 2010 Elastic waves and transition to elastic
  turbuelnce in a two-dimensional viscoelastic {K}olmogorov flow. {\em Phys.
  Rev. E\/} {\bf 82}, 036314.

\bibitem[Boffetta {\em et~al.\/}(2005)Boffetta, Celani, Mazzino, Puliafito \&
  Vergassola]{Boffetta05}
{\sc Boffetta, G., Celani, A., Mazzino, A., Puliafito, A. \& Vergassola, M.}
  2005 The viscoelastic {K}olmogorov flow: eddy viscosity and linear
  instability. {\em J. Fluid Mech.\/} {\bf 523}, 161--170.

\bibitem[Buza {\em et~al.\/}(2022{\natexlab{{\em a\/}}})Buza, Beneitez, Page \&
  Kerswell]{Buza22b}
{\sc Buza, Gergely, Beneitez, Miguel, Page, Jacob \& Kerswell, Rich~R}
  2022{\natexlab{{\em a\/}}} Finite-amplitude elastic waves in viscoelastic
  channel flow from large to zero {R}eynolds number. {\em Journal of Fluid
  Mechanics\/} {\bf 951}, A3.

\bibitem[Buza {\em et~al.\/}(2022{\natexlab{{\em b\/}}})Buza, Page \&
  Kerswell]{Buza22a}
{\sc Buza, Gergely, Page, Jacob \& Kerswell, Rich~R} 2022{\natexlab{{\em b\/}}}
  Weakly nonlinear analysis of the viscoelastic instability in channel flow for
  finite and vanishing {R}eynolds numbers. {\em Journal of Fluid Mechanics\/}
  {\bf 940}, A11.

\bibitem[Chaudhary {\em et~al.\/}(2021)Chaudhary, Garg, Subramanian \&
  Shankar]{Chaudhary21}
{\sc Chaudhary, Indresh, Garg, P., Subramanian, Ganesh \& Shankar, V.} 2021
  Linear instability of viscoelastic pipe flow. {\em Journal of Fluid
  Mechanics\/} {\bf 908}, A11.

\bibitem[Couchman {\em et~al.\/}(2023)Couchman, Beneitez, Page \&
  Kerswell]{Couchman23}
{\sc Couchman, M. M.~P., Beneitez, M., Page, J. \& Kerswell, R.~R.} 2023
  Inertial enhancement of the polymer diffusive instability. {\em
  http://arxiv.org/abs/2308.14879\/} .

\bibitem[Datta {\em et~al.\/}(2022)Datta, Ardekani, Arratia, Beris,
  Bischofberger, McKinley, Eggers, L{\'o}pez-Aguilar, Fielding, Frishman {\em
  et~al.\/}]{Datta22}
{\sc Datta, Sujit~S, Ardekani, Arezoo~M, Arratia, Paulo~E, Beris, Antony~N,
  Bischofberger, Irmgard, McKinley, Gareth~H, Eggers, Jens~G,
  L{\'o}pez-Aguilar, J~Esteban, Fielding, Suzanne~M, Frishman, Anna {\em
  et~al.\/}} 2022 Perspectives on viscoelastic flow instabilities and elastic
  turbulence. {\em Physical Review Fluids\/} {\bf 7}~(8), 080701.

\bibitem[Dong \& Zhang(2022)]{Dong22}
{\sc Dong, M. \& Zhang, M.} 2022 Asymptotic study of linear instability in a
  viscoelastic pipe flow. {\em Journal of Fluid Mechanics\/} {\bf 935}, A28.

\bibitem[Drazin \& Reid(1981)]{DrazinReid}
{\sc Drazin, P.~G. \& Reid, W.~H.} 1981 {\em Hydrodynamic Stability\/} .

\bibitem[Dubief {\em et~al.\/}(2022)Dubief, Page, Kerswell, Terrapon \&
  Steinberg]{Dubief22}
{\sc Dubief, Yves, Page, Jacob, Kerswell, Richard~R, Terrapon, Vincent~E \&
  Steinberg, Victor} 2022 First coherent structure in elasto-inertial
  turbulence. {\em Physical Review Fluids\/} {\bf 7}~(7), 073301.

\bibitem[Dubief {\em et~al.\/}(2023)Dubief, Terrapon \& Hof]{Dubief23}
{\sc Dubief, Y., Terrapon, V.~E. \& Hof, B.} 2023 Elasto-inertial turbulence.
  {\em Annual Review of Fluid Mechanics\/} {\bf 55}, 675--705.

\bibitem[Garg {\em et~al.\/}(2018)Garg, Chaudhary, Khalid, Shankar \&
  Subramanian]{Garg18}
{\sc Garg, Piyush, Chaudhary, Indresh, Khalid, Mohammad, Shankar, V \&
  Subramanian, Ganesh} 2018 Viscoelastic pipe flow is linearly unstable. {\em
  Physical Review Letters\/} {\bf 121}~(2), 024502.

\bibitem[Groisman \& Steinberg(2000)]{groisman00}
{\sc Groisman, Alexander \& Steinberg, Victor} 2000 Elastic turbulence in a
  polymer solution flow. {\em Nature\/} {\bf 405}~(6782), 53--55.

\bibitem[Khalid {\em et~al.\/}(2021{\natexlab{{\em a\/}}})Khalid, Chaudhary,
  Garg, Shankar \& Subramanian]{Khalid21a}
{\sc Khalid, Mohammad, Chaudhary, Indresh, Garg, Piyush, Shankar, V \&
  Subramanian, Ganesh} 2021{\natexlab{{\em a\/}}} The centre-mode instability
  of viscoelastic plane {P}oiseuille flow. {\em Journal of Fluid Mechanics\/}
  {\bf 915}.

\bibitem[Khalid {\em et~al.\/}(2021{\natexlab{{\em b\/}}})Khalid, Shankar \&
  Subramanian]{Khalid21b}
{\sc Khalid, Mohammad, Shankar, V \& Subramanian, Ganesh} 2021{\natexlab{{\em
  b\/}}} Continuous pathway between the elasto-inertial and elastic turbulent
  states in viscoelastic channel flow. {\em Physical Review Letters\/} {\bf
  127}~(13), 134502.

\bibitem[Larson {\em et~al.\/}(1990)Larson, Shaqfeh \& Muller]{Larson90}
{\sc Larson, Ronald~G, Shaqfeh, Eric~SG \& Muller, Susan~J} 1990 A purely
  elastic instability in {T}aylor--{C}ouette flow. {\em Journal of Fluid
  Mechanics\/} {\bf 218}, 573--600.

\bibitem[Lellep {\em et~al.\/}(2023{\natexlab{{\em a\/}}})Lellep, Linkmann \&
  Morozov]{Lellep23a}
{\sc Lellep, M., Linkmann, M. \& Morozov, A.} 2023{\natexlab{{\em a\/}}} Linear
  instability analysis of purely elastic travelling waves in pressure-driven
  channel flows. {\em J. Fluid Mech.\/} {\bf 959}, R1.

\bibitem[Lellep {\em et~al.\/}(2023{\natexlab{{\em b\/}}})Lellep, Linkmann \&
  Morozov]{Lellep23b}
{\sc Lellep, M., Linkmann, M. \& Morozov, A.} 2023{\natexlab{{\em b\/}}} Purely
  elastic turbulence in pressure-driven channel flows. {\em
  http://arxiv.org/arXiv.2312.0891\/} .

\bibitem[Lewy \& Kerswell(2023)]{Lewy23}
{\sc Lewy, T. \& Kerswell, R.~R.} 2023 The polymer diffusive instability in
  highly concentrated polymer fluids. {\em https://arxiv.org/abs/2311.05251\/}
  .

\bibitem[Morozov(2022)]{Morozov22}
{\sc Morozov, Alexander} 2022 Coherent structures in plane channel flow of
  dilute polymer solutions with vanishing inertia. {\em Physical Review
  Letters\/} {\bf 129}~(1), 017801.

\bibitem[Page {\em et~al.\/}(2020)Page, Dubief \& Kerswell]{Page20}
{\sc Page, Jacob, Dubief, Yves \& Kerswell, Rich~R} 2020 Exact traveling wave
  solutions in viscoelastic channel flow. {\em Physical Review Letters\/} {\bf
  125}~(15), 154501.

\bibitem[Rallison \& Hinch(1995)]{Rallison95}
{\sc Rallison, J.~M. \& Hinch, E.~J.} 1995 Instability of a high-speed
  submerged elastic jet. {\em J. Fluid Mech.\/} {\bf 228}, 311--324.

\bibitem[Samanta {\em et~al.\/}(2013)Samanta, Dubief, Holzner, Sch{\"a}fer,
  Morozov, Wagner \& Hof]{samanta13}
{\sc Samanta, Devranjan, Dubief, Yves, Holzner, Markus, Sch{\"a}fer, Christof,
  Morozov, Alexander~N, Wagner, Christian \& Hof, Bj{\"o}rn} 2013
  Elasto-inertial turbulence. {\em Proceedings of the National Academy of
  Sciences\/} {\bf 110}~(26), 10557--10562.

\bibitem[Shaqfeh(1996)]{Shaqfeh96}
{\sc Shaqfeh, ES~G} 1996 Purely elastic instabilities in viscometric flows.
  {\em Annual Review of Fluid Mechanics\/} ~(28), 129--185.

\bibitem[Shekar {\em et~al.\/}(2021)Shekar, McMullen, McKeon \&
  Graham]{shekar21}
{\sc Shekar, A., McMullen, R.~M., McKeon, B.~J. \& Graham, M.~D.} 2021
  Tollmien-{S}chlichting route to elastoinertial turbulence in channel flow.
  {\em Physical Review Fluids\/} {\bf 6}, 093301.

\bibitem[Shekar {\em et~al.\/}(2019)Shekar, McMullen, Wang, McKeon \&
  Graham]{shekar19}
{\sc Shekar, A., McMullen, R.~M., Wang, S-N., McKeon, B.~J. \& Graham, M.~D.}
  2019 Critical-layer structures and mechanisms in elastoinertial turbulence.
  {\em Phys. Rev. Lett.\/} {\bf 122}, 124503.

\bibitem[Sid {\em et~al.\/}(2018)Sid, Terrapon \& Dubief]{Sid18}
{\sc Sid, Samir, Terrapon, VE \& Dubief, Y} 2018 Two-dimensional dynamics of
  elasto-inertial turbulence and its role in polymer drag reduction. {\em
  Physical Review Fluids\/} {\bf 3}~(1), 011301.

\bibitem[Wan {\em et~al.\/}(2021)Wan, Sun \& Zhang]{Wan21}
{\sc Wan, D., Sun, G. \& Zhang, M.} 2021 Subcritical and suprcritical
  bifurcations in axisymmetric viscoelastic pipe flows. {\em Journal of Fluid
  Mechanics\/} {\bf 929}, A16.

\end{thebibliography}
\end{document}